\documentclass[10pt,aps,pra,twocolumn,superscriptaddress,longbibliography,floatfix]{revtex4-2}
\usepackage{array}[=2016-10-06]
\usepackage{amsmath,amssymb,amsfonts}
\usepackage{bm}
\usepackage{braket}
\usepackage{mathrsfs}
\usepackage{siunitx}
\usepackage{graphicx}
\usepackage{dcolumn}
\usepackage{booktabs}
\usepackage{tabularx,makecell,adjustbox}
\newcolumntype{Y}{>{\raggedright\arraybackslash}X}
\newcolumntype{C}{>{\centering\arraybackslash}X}
\usepackage{silence}
\usepackage{placeins}
\usepackage{caption}
\usepackage{subcaption}
\usepackage{xcolor}
\definecolor{RevisionRed}{RGB}{0,0,0}
\definecolor{RevisionBlue}{RGB}{0,0,0}
\newcommand{\revred}[1]{\begingroup\color{RevisionRed}#1\endgroup}
\newenvironment{redrevision}{\begingroup\color{RevisionRed}}{\endgroup}

\usepackage{comment}
\usepackage{silence}
\usepackage{ragged2e}
\usepackage{fancyhdr}
\usepackage{orcidlink}
\hypersetup{
  colorlinks=true,
  linkcolor=red,
  citecolor=blue,
  urlcolor=blue
}

\begin{document}
\title{Chip-integrated superconducting gravimetric sensing via flux-mediated transmon mechanical coupling: Theory and device requirements}
\author{Salman Sajad Wani\,\orcidlink{0000-0002-5262-9738}}
\email{sawa54922@hbku.edu.qa}
\affiliation{Qatar Center for Quantum Computing, College of Science and Engineering, Hamad Bin Khalifa University, Doha, Qatar}
\author{Mughees Ahmed Khan\,\orcidlink{0009-0009-4214-7637}}
\email{mukh68937@hbku.edu.qa}
\affiliation{Qatar Center for Quantum Computing, College of Science and Engineering, Hamad Bin Khalifa University, Doha, Qatar}
\author{Abrar Ahmed Naqash\,\orcidlink{0000-0003-3891-4740}}
\email{abrar.naqash@candqrc.ca}
\affiliation{Canadian Quantum Research Center, 204-3002 32 Ave, Vernon, BC V1T 2L7, Canada}
\author{Saif Al-Kuwari\,\orcidlink{0000-0002-4402-7710}}
\affiliation{Qatar Center for Quantum Computing, College of Science and Engineering, Hamad Bin Khalifa University, Doha, Qatar}
\begin{abstract}
Measurements of local gravitational acceleration support geophysics and inertial navigation. We develop a circuit-level model of a chip-integrated superconducting gravimetric sensing head that longitudinally couples a nanomechanical beam in a superconducting quantum interference device (SQUID) loop to an independently flux-biased transmon. \revred{The interaction encodes changes in gravitational acceleration in the qubit Ramsey phase.} We derive the phase response and evaluate the quantum Fisher information of the reduced qubit state and the classical Fisher information accessible through Ramsey readout. The model includes qubit relaxation and dephasing, mechanical damping, flux noise, readout infidelity, and cycle overhead. \revred{With other model inputs fixed, the competition between phase gain and responsivity-dependent decoherence yields a finite optimal transmon frequency responsivity to mechanical-loop flux.} The decoherence model includes dephasing from low-frequency flux noise in the mechanical SQUID loop and coupling-dependent qubit dephasing from the mechanical bath. For the prospective integrated-device parameter set, Ramsey readout at the first mechanical revival gives a conditional projection of the sensing head's sensitivity of $1.1\times10^{-5}\,\mathrm{m\,s^{-2}\,Hz^{-1/2}}$ at an interrogation time of $10.0\,\mu\mathrm{s}$. This estimate uses the single-cycle-equivalent Fisher-information rate for statistically independent cycles. A separate analysis identifies the joint mechanical, circuit, flux-noise, coherence, timing, and readout parameters needed within the model to reach $10^{-6}$-class internal sensitivity. \revred{These results specify the design tradeoffs and component requirements for Ramsey gravimetry in the proposed SQUID--transmon electromechanical sensing head.}
\end{abstract}
\maketitle
\noindent{\small\itshape This article may be downloaded for personal use only. Any other use requires prior permission of the author and AIP Publishing. This article appeared in AVS Quantum Sci.\ \textbf{8} (2026) and may be found at \url{https://doi.org/10.1116/5.0356589}.\par}
\vspace{4pt}
\section{Introduction}

High-precision measurements of local gravitational acceleration support tests of fundamental physics \cite{PhysRevD.55.455}, geophysical monitoring \cite{Bidel2018,AntoniMicollier2022}, subsurface mapping \cite{Stray2022}, and inertial navigation \cite{Bidel2018}. Free-fall corner-cube gravimeters \cite{Niebauer1995RSI} and cold-atom gravimeters \cite{Peters1999Nature,PhysRevLett.106.038501} provide absolute measurements of gravitational acceleration. Spring-based relative gravimeters provide continuous measurements whose performance depends on calibration, drift correction, and environmental control \cite{Crossley2013RG}. MEMS gravimeters extend this approach to compact devices, with performance governed by drift, environmental coupling, and the bandwidth--sensitivity trade-off \cite{Wu2025,Gao2026}. Superconducting gravimeters achieve high stability at low frequencies \cite{Prothero1968RSI,Crossley2013RG,Hinderer2022} and are used to measure Earth tides and Earth's normal modes \cite{Crossley2013RG}. These instruments require dedicated cryogenic instrumentation \cite{Crossley2013RG,Hinderer2022}. Within a dilution-refrigerator environment appropriate for a transmon circuit, we analyze a chip-integrated sensing element based on a superconducting qubit. We determine how circuit constraints and decoherence affect the encoding of residual acceleration in the qubit readout signal.

Gravimetric protocols based on conditional mechanical displacements have been studied in spin--mechanical systems \cite{ChenYin2018,Montenegro2025}. In these protocols, the mechanical trajectories associated with the two spin states close in phase space, leaving a force-dependent phase for readout. A longitudinal qubit--oscillator sensing scheme uses a closed phase-space trajectory to generate a geometric phase measured through the qubit \cite{Suri2025}. Optomechanical architectures have also been proposed for quantum gravimetry \cite{Armata2017,Qvarfort2018}, together with a superconducting {magnetomechanical} scheme \cite{Johnsson2016}. {A distinct transmon-based proposal encodes gravity through gravitational redshift and the gravitational Aharonov--Bohm phase~\cite{Balatsky2025}.} SQUID-based displacement transduction has been demonstrated experimentally \cite{Etaki2008NatPhys}, and flux-mediated longitudinal coupling between a transmon and a mechanical resonator has been demonstrated experimentally \cite{bera2021}. {The sensing architecture studied here combines these previously developed ingredients.} {Prior electromechanical studies relate transmon flux responsivity to longitudinal coupling~\cite{kounalakis,bera2021}.} {Transmon-coherence measurements show how the same responsivity converts low-frequency flux noise into dephasing~\cite{Hutchings2017}.} In this two-loop architecture, the transmon flux responsivity is the change in transition frequency per unit flux. {The circuit-level question addressed here is how the mechanical-loop responsivity jointly determines gravimetric phase gain, responsivity-dependent decoherence, Fisher-information rate, and integrated-device requirements.}

We develop a circuit-level model of this chip-integrated sensing element for operation within a cryogenic instrument. Following the flux-mediated SQUID--transmon architecture proposed by Kounalakis et al. \cite{kounalakis}, a nanomechanical beam forms part of one SQUID loop, termed the mechanical SQUID. This loop is connected in parallel with a second, independently flux-biased SQUID that controls the transmon transition frequency. Beam motion changes the flux through the mechanical SQUID, thereby shifting the transmon transition frequency. This modulation produces an effective longitudinal interaction between the qubit and the mechanical mode.

\revred{We describe acceleration along the sensing axis relative to a fixed baseline. The residual acceleration enters the model as a force on the mechanical mode.} \revred{The gravitational loading and longitudinal interaction produce an acceleration-dependent qubit phase.} After one mechanical period, the conditional trajectories close in phase space and the qubit and resonator ideally disentangle; a Ramsey measurement then reads out the accumulated qubit phase. In the lossless limit, separation of the conditional mechanical states temporarily reduces the qubit coherence, with a larger reduction at higher thermal occupation. Phase-space closure restores this coherence at the revival. We evaluate the protocol in the presence of mechanical damping, flux-noise dephasing, qubit energy relaxation, intrinsic pure dephasing, and finite readout fidelity.

{Using the established flux-mediated coupling relation for transmon electromechanics~\cite{kounalakis,bera2021}, we express the Pauli longitudinal coupling $g_z$ in terms of the mechanical-loop responsivity $D_m$.} {We then calculate the reduced-qubit quantum Fisher information and show that an optimally phased ideal Ramsey quadrature measurement attains it within the phase-encoding model.} {At the first mechanical revival, with all other model parameters fixed, the phase slope grows linearly with $D_m$.} {Under the same conditions, the same-loop flux-noise exponent $\chi_\Phi$ and the mechanical-bath exponent $\chi_m$ grow quadratically.} {The competition between phase gain and these response-dependent losses yields a finite optimum.} We evaluate two parameter sets: one assembled from values reported separately for relevant device components and one representing a prospective integrated device. {We report conditional sensing-head sensitivities using the single-cycle-equivalent Fisher-information rate for statistically independent cycles.} {Each one-period interrogation ends at the first mechanical revival.} The parameter set assembled from separately reported component values gives a projected acceleration sensitivity of approximately $1.6\times10^{-5}\,\mathrm{m\,s^{-2}\,Hz^{-1/2}}$. The prospective integrated-device parameter set gives approximately $1.1\times10^{-5}\,\mathrm{m\,s^{-2}\,Hz^{-1/2}}$. {A separate requirements analysis identifies a rounded $12~\mathrm{kHz}$ mechanical-frequency point under a specified low-frequency flux-noise assumption. At this point, the model projects a conditional internal sensitivity of $1.0\times10^{-6}\,\mathrm{m\,s^{-2}\,Hz^{-1/2}}$ for the proposed sensor head and specifies the associated joint mechanical, circuit, flux-noise, coherence, timing, and readout requirements.} \revred{The sensitivity estimates apply to local estimation of residual acceleration about a fixed reference operating point.} Achieving instrument-level absolute accuracy and long-term stability requires device-specific calibration, measured transfer functions for vibration and tilt coupling, and characterization of temperature-induced flux drift and time-correlated noise.

\section{\texorpdfstring{{Hybrid SQUID--Transmon Electromechanical Platform}}{Hybrid SQUID--Transmon Electromechanical Platform}}
\label{sec2}

\begin{figure}
    \centering
    \includegraphics[width=\columnwidth]{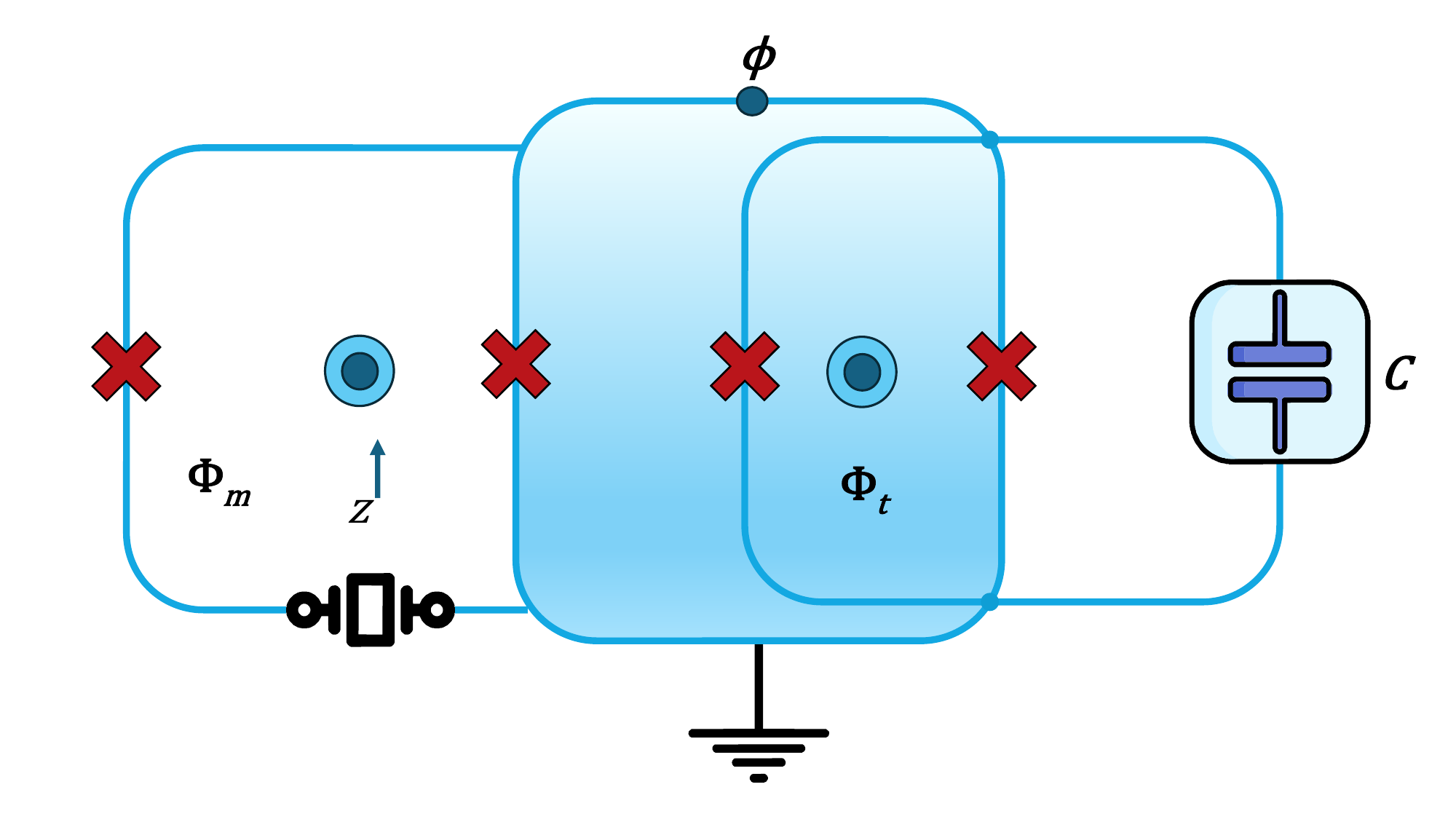}
    \caption{\justifying Schematic of the hybrid SQUID--transmon {gravimetric sensing head}. A nanomechanical oscillator (vertical displacement $z$) is embedded in a mechanical SQUID loop threaded by flux $\Phi_m$. This loop is connected in parallel to a flux-tunable transmon SQUID threaded by flux $\Phi_t$. Red crosses indicate Josephson junctions, with two junctions in each SQUID loop, and $C$ is the transmon shunt capacitance. {$\phi$ denotes the circuit node flux relative to ground.}}
    \label{fig:squida}
\end{figure}

\begin{figure*}
    \centering
    \includegraphics[width=\textwidth]{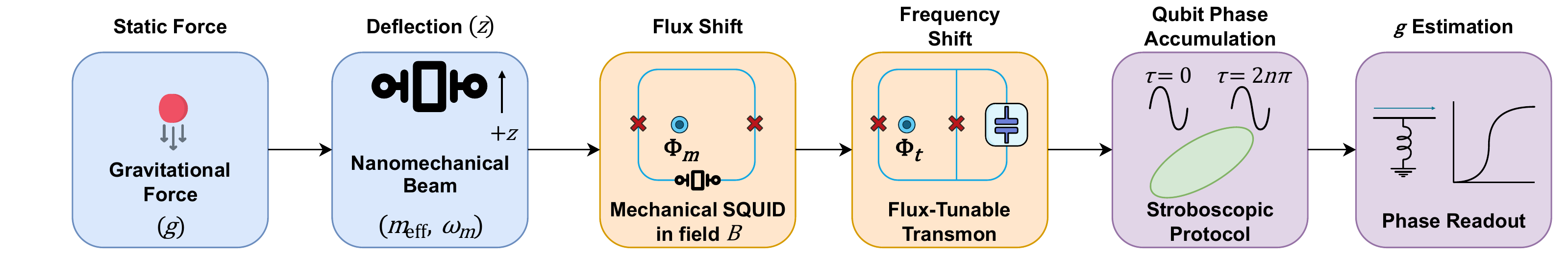}
    \caption{\justifying Operational principle of the hybrid SQUID--transmon {gravimetric sensing head}. \revred{Residual gravitational acceleration $g$ changes the beam displacement and hence the qubit frequency through the longitudinal interaction.} \revred{Ramsey readout near a mechanical revival measures the resulting qubit phase relative to the calibrated baseline.}}
    \label{fig:squidb}
\end{figure*}

{We model a proposed chip-integrated gravimetric sensing head comprising a flux-tunable transmon and a high-$Q$ nanomechanical beam.} Following the architecture of Kounalakis \emph{et al.}~\cite{kounalakis}, the device consists of two superconducting loops connected in parallel {(Fig.~\ref{fig:squida})}, a mechanical SQUID, in which the suspended beam is embedded, biased by flux $\Phi_m$, and a flux-tunable transmon SQUID biased by $\Phi_t$. With an in-plane magnetic field $B$, {out-of-plane} beam motion modulates the mechanical-SQUID flux and hence the transmon frequency $\Omega_q(\Phi_m,\Phi_t)$, yielding a longitudinal ($\propto \sigma_z$) qubit--mechanics coupling. Figure~\ref{fig:squidb} outlines the sensing sequence.

\revred{A beam displacement $z$ {from the fixed reference position} produces a SQUID-flux shift $\delta\Phi(z)=\lambda B\ell z$, where $\lambda$ is the mode-shape factor and $\ell$ is the lever arm.} \revred{We denote the physical acceleration along the sensing axis by $g_{\rm phys}$, choose a fixed baseline $g_{\rm ref}$, and use $g\equiv\delta g=g_{\rm phys}-g_{\rm ref}$ for the residual acceleration throughout the model.} \revred{We take $g_{\rm phys}$ and $g_{\rm ref}$ as positive downwards and mechanical displacement as positive upwards.} \revred{We write the dimensionless interrogation time as $\tau=\omega_m t$ and use $\sigma_z|0\rangle=-|0\rangle$ and $\sigma_z|1\rangle=|1\rangle$.}

\revred{To define the reference explicitly, let ${z_{\rm tot}}$ be the upward mechanical displacement before shifting the origin and $K=m_{\rm eff}\omega_m^2$ the stiffness.} \revred{Within the linear operating interval, write the qubit angular frequency as $\Omega({z_{\rm tot}})=\Omega_0+{S_z}{z_{\rm tot}}$, with signed displacement slope ${S_z}=\partial\Omega/\partial {z_{\rm tot}}$.} \revred{Adopting the number-operator coupling convention used by Kounalakis \emph{et al.}~\cite{kounalakis}, we write the effective qubit energy as $H_q({z_{\rm tot}})=\hbar\Omega({z_{\rm tot}})c^\dagger c$, where $c^\dagger c$ is the transmon excitation-number operator.} In the qubit subspace, $c^\dagger c=(I+\sigma_z)/2$. \revred{The identity component contains the common electrical energy $\overline\lambda {z_{\rm tot}}$, with $\overline\lambda=\hbar {S_z}/2$, and hence produces the qubit-independent force $-\overline\lambda$.} \revred{The common potential before shifting the origin is $U_{\rm com}({z_{\rm tot}})=K{z_{\rm tot}}^2/2+ (\overline\lambda+m_{\rm eff}g_{\rm phys}){z_{\rm tot}}$.}

\revred{The fixed reference position, displacement coordinate and qubit frequency are}
\begin{redrevision}
\begin{equation}
\begin{aligned}
{z_{\rm ref}}
&=-\frac{\overline\lambda+m_{\rm eff}g_{\rm ref}}{K},
\qquad {z}={z_{\rm tot}}-{z_{\rm ref}},\\
\Omega_q&=\Omega_0+{S_z}{z_{\rm ref}}.
\end{aligned}
\label{eq:fixed-reference-definitions}
\end{equation}
\end{redrevision}
\revred{The common potential then becomes}
\begin{redrevision}
\begin{equation}
U_{\rm com}({z})
=\frac{K}{2}{z}^2+m_{\rm eff}g\,{z}+C(g),
\label{eq:shifted-common-potential}
\end{equation}
\end{redrevision}
\revred{where $C(g)$ contributes only an overall phase.} \revred{The reference position accounts for the baseline loading and common electrical force and is held fixed throughout a measurement series. No active force-balancing loop, compensation-force modulation or shot-by-shot change of origin is assumed.}

\revred{We quantize the displacement as ${z}=z_{\rm zpf}(a+a^\dagger)$, where $a$ is the mechanical annihilation operator and $z_{\rm zpf}=\sqrt{\hbar/(2m_{\rm eff}\omega_m)}$.} \revred{The number-operator coupling rate is $g_n={S_z}z_{\rm zpf}$, so $\overline\lambda=\hbar g_n/(2z_{\rm zpf})$.} \revred{The Pauli longitudinal coupling rate is $g_z=g_n/2$, while $\mathcal{G}=m_{\rm eff}g z_{\rm zpf}/\hbar$ is the gravitational drive rate.} \revred{Omitting scalar energy terms, the system Hamiltonian in the longitudinal coupling regime is}
\begin{equation}
\label{eq:H-longitudinal}
\frac{\mathcal{H}}{\hbar}
 = \frac{\Omega_q}{2}\sigma_z + \omega_m a^\dagger a
 + g_z\,\sigma_z(a+a^\dagger)
 + \mathcal{G}\,(a+a^\dagger).
\end{equation}
\revred{Local derivatives with respect to the residual acceleration $g$ are taken with $g_{\rm ref}$, the circuit biases, $\Omega_q$, $\omega_m$ and coupling parameters fixed.} \revred{The relation between flux responsivity and coupling demonstrated by Bera \emph{et al.}~\cite{bera2021} gives}
\begin{redrevision}
\begin{equation}
\frac{|g_n|}{2\pi}
=D_m|\delta\varphi_{\rm zpf}|,
\qquad
\delta\varphi_{\rm zpf}
=\frac{\lambda B\ell z_{\rm zpf}}{\Phi_0}.
\end{equation}
\end{redrevision}
where $D_m=\left|\partial f_q/\partial(\Phi_m/\Phi_0)\right|$ is the frequency response to the mechanical-loop flux. \revred{With $f_m=\omega_m/(2\pi)$, the signed dimensionless coupling $k$ and its magnitude satisfy}
\begin{redrevision}
\begin{equation}
\begin{aligned}
g_z&=\frac{g_n}{2},
\qquad k=\frac{g_z}{\omega_m},\\
|k|&=\frac{|g_z|}{\omega_m}
=\frac{D_m|\delta\varphi_{\rm zpf}|}{2f_m}.
\end{aligned}
\label{eq:coupling-map}
\end{equation}
\end{redrevision}
\revred{All tabulated coupling strengths are magnitudes.}

The two-loop derivative $D_m$ follows the independently biased mechanical and transmon SQUID coordinates of Kounalakis \emph{et al.}~\cite{kounalakis}. Scenario~I is a composite component benchmark. It combines the demonstrated $40.9$-kHz, $4$-ng mechanical scale~\cite{Reinhardt2016} with field, lever-arm, linewidth, and coupling scales from integrated flux electromechanics~\cite{Bera2024}. In Bera \emph{et al.}, the approximately $6~\mathrm{Hz}$ linewidth, corresponding to $Q_m\simeq6.6\times10^5$ for the $3.97$-MHz mode, was measured near $18~\mathrm{mT}$; the approximately $27~\mathrm{mT}$ coupling point had a linewidth near $13~\mathrm{Hz}$. The $20~\mathrm{mK}$ entry used here is a refrigerator-plate-temperature assumption. Bera \emph{et al.} inferred a mechanical-mode temperature near $70~\mathrm{mK}$. These entries are component-level benchmark inputs drawn from separate operating conditions. For the dimensionless flux $\varphi=\Phi_m/\Phi_0$, we use $\sqrt{A_\varphi}=1.4\times10^{-6}$ in the $1/f$ spectrum defined explicitly below. Under the convention in Eq.~\eqref{eq:flux-psd}, this coefficient corresponds to the one-sided one-hertz amplitude spectral density $\sqrt{S_\Phi^{(1)}(1~\mathrm{Hz})} =1.98~\mu\Phi_0/\sqrt{\mathrm{Hz}}$~\cite{Hutchings2017}. Scenario~II adopts a $100$-kHz, $530$-ng, $100$-mT, $50$-$\mu$m mechanical core as a long-term integration target and uses $Q_m=10^6$; the field and lever-arm scales have been demonstrated in dc-SQUID mechanical transduction~\cite{Poot2010}. The listed values of $D_m$ minimize the one-period sensitivity within Eq.~\eqref{eq:visibility-model}; $D_m$ and the background Ramsey coherence are explicit integrated-device requirements.

\revred{We define the scaled gravity $\bar{G}=\mathcal{G}/\omega_m$, leading to the operator-valued displacement amplitude $\mathcal{Z}=k\sigma_z+\bar{G}$.} \revred{This yields the scaled Hamiltonian also given in Appendix~\ref{AA}, Eq.~\eqref{s3}:}
\begin{equation}
\mathcal{H}
= \hbar\omega_m
\bigl[a^\dagger a+\mathcal{Z}(a+a^\dagger)\bigr]
+\frac{\hbar\Omega_q}{2}\sigma_z.
\label{eq:Hdimless}
\end{equation}

\begin{table}[t]
\caption{\justifying \textbf{Scenario-I device parameters.} Component-anchored values used for the principal conditional {sensing-head} estimate~\cite{Reinhardt2016,Bera2024,Hutchings2017}.}
\label{tab:params-main}
\centering
\small
\renewcommand{\arraystretch}{1.2}
\setlength{\tabcolsep}{6pt}
\begin{tabular}{l c}
\hline\hline
\textbf{Quantity} & \textbf{Value} \\
\hline
Mechanical frequency & $\omega_m/2\pi=40.9~\mathrm{kHz}$ \\
Effective mass & $m_{\rm eff}=4.0~\mathrm{ng}$ \\
Magnetic field & $B=27~\mathrm{mT}$ \\
Lever arm & $\ell=40~\mu\mathrm{m}$ \\
Mode-shape factor & $\lambda=0.9$ \\
Mechanical-loop response & \revred{$D_m=0.868~\mathrm{GHz}$} \\
Number-operator coupling & $g_n/2\pi=2.923~\mathrm{kHz}$ \\
Pauli coupling & $g_z/2\pi=1.462~\mathrm{kHz}$ \\
Coupling ratio & $k=0.03574$ \\
Mechanical quality factor & $Q_m=6.6\times10^5$ \\
Bath temperature & $T=20~\mathrm{mK}$ \\
Normalized $1/f$ amplitude & $\sqrt{A_\varphi}=1.4\times10^{-6}$ \\
Background Ramsey time & $T_{2R,{\rm bg}}=50~\mu\mathrm{s}$ \\
\hline\hline
\end{tabular}
\end{table}

\revred{We retain the linear Hamiltonian over the residual operating interval about the fixed reference position.} An integrated device must verify this interval by measuring the curvature $\partial^2 f_q/\partial(\Phi_m/\Phi_0)^2$. Applying a unitary polaron transformation factorizes the time-evolution operator Eq.~\eqref{s4}:
\begin{align}
\label{eq:evolution}
\mathcal{U}(\tau)
={}&
\exp\left(-i\frac{\Omega_q}{2\omega_m}\tau\sigma_z\right)
\exp\left[i\mathcal{Z}^2(\tau-\sin\tau)\right]
\nonumber\\
&\times
\exp\left[
-\mathcal{Z}
\left((1-e^{-i\tau})a^\dagger-(1-e^{i\tau})a\right)
\right]
\nonumber\\
&\times\exp\left(-i\tau a^\dagger a\right).
\end{align}
\revred{The factor $\exp[i\mathcal{Z}^2(\tau-\sin\tau)]$ contributes a gravity-dependent qubit phase.} \revred{The complete Ramsey phase can also depend on the mechanical input, but this dependence vanishes at ideal exact revivals $\tau=2\pi n$.} \revred{The locally fixed preparation, held fixed as $g$ varies, and the statically loaded preparation are treated separately below.}

\revred{A baseline calibration precedes the measurement series with the device orientation, magnetic field and SQUID flux biases fixed.} \revred{The baseline Ramsey measurement sets the reference phase, while the calibrated mechanical and circuit parameters determine the signed acceleration-to-phase response.}
\revred{For the loaded preparation analyzed in Sec.~\ref{sec:loaded-preparation}, gravity and the longitudinal coupling are present while the qubit is initialized and held in $|0\rangle$ and the mechanics settles to the displaced thermal state.} \revred{Initial settling occurs before data acquisition; a change in gravitational loading requires settling again or explicit treatment of the mechanical transient.} \revred{A short microwave $\pi/2$ pulse prepares a balanced qubit superposition without changing the mechanical state. Free interrogation under Eq.~\eqref{eq:H-longitudinal} follows.} \revred{For the reported benchmarks, interrogation ends at the first nominal revival $t_{\rm rev}=2\pi/\omega_m$; a second $\pi/2$ pulse with controlled phase converts the qubit phase into a population difference for binary readout, after which the qubit is reset.} \revred{Initialization and both pulses are modeled as ideal operations, with coupled evolution during the pulses neglected; the complete-cycle conditions are stated in Sec.~\ref{sec:addnotes}.}
\revred{The measured residual can be constant during interrogation, so the procedure supports relative quasi-static sensing.} \revred{If $g_{\rm ref}$ is independently known, $g_{\rm phys}=g_{\rm ref}+g$ can be reconstructed with the reference, baseline-phase and response-calibration uncertainties propagated.}

\section{Quantum Fisher Information and Sensitivity Analysis}
\label{sec3}

\begin{figure*}[t!]
    \centering
    \begin{subfigure}[b]{0.32\textwidth}
        \centering
        \includegraphics[width=\textwidth]{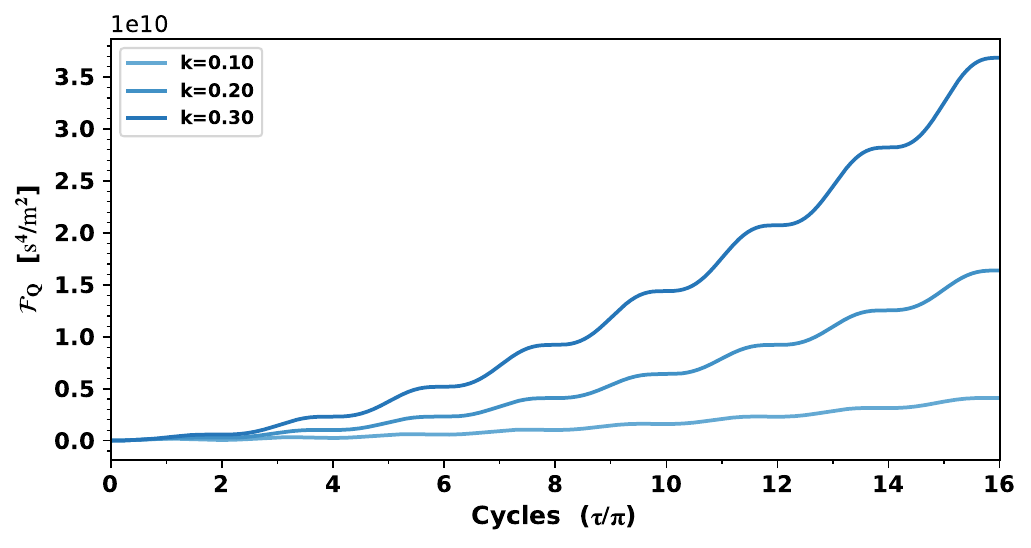}
        \caption{}
        \label{fig:qfi_linear_panel_a}
    \end{subfigure}\hfill
    \begin{subfigure}[b]{0.32\textwidth}
        \centering
        \includegraphics[width=\textwidth]{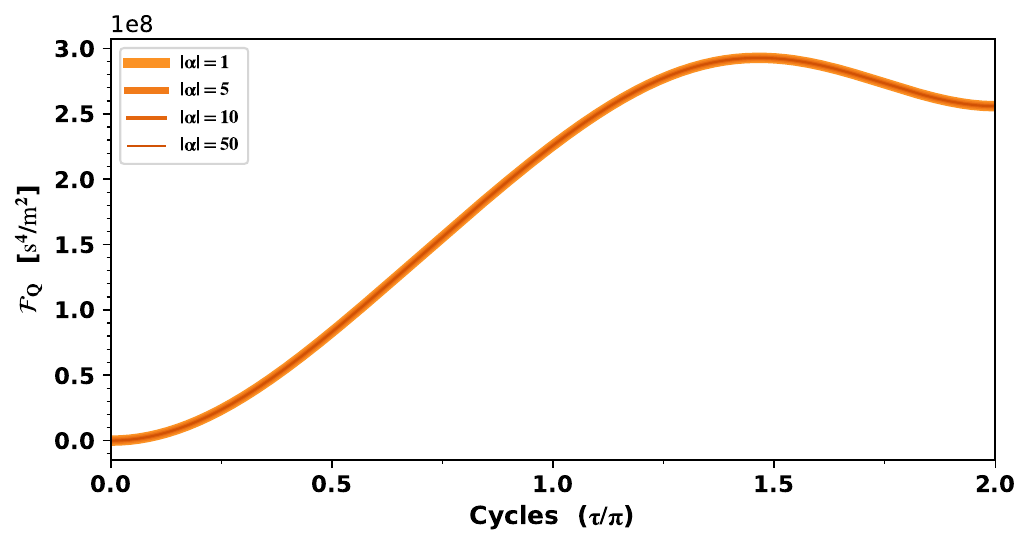}
        \caption{}
        \label{fig:qfi_linear_panel_b}
    \end{subfigure}\hfill
    \begin{subfigure}[b]{0.32\textwidth}
        \centering
        \includegraphics[width=\textwidth]{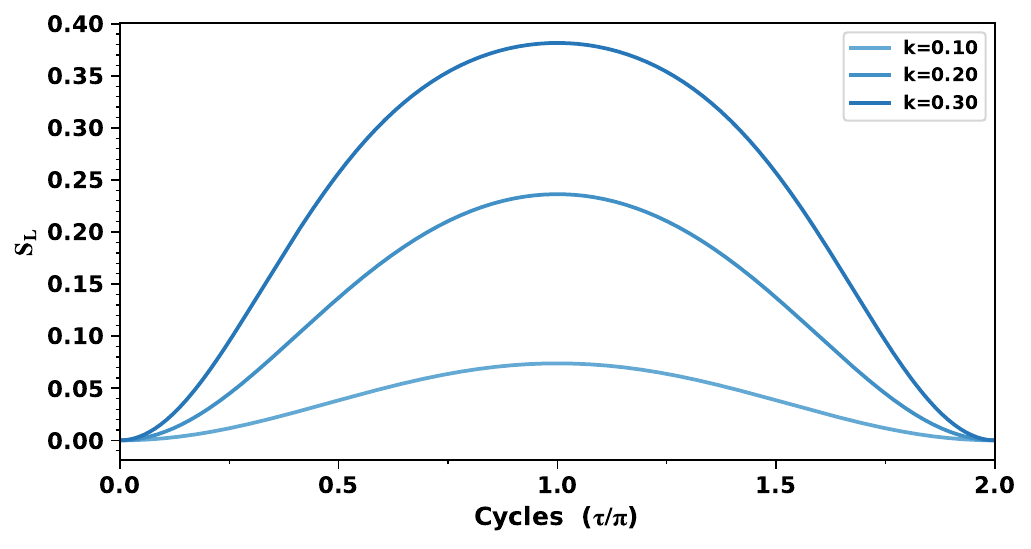}
        \caption{}
        \label{fig:qfi_linear_panel_c}
    \end{subfigure}
    \caption{\justifying Quantum Fisher information (QFI) and linear entropy as functions of {dimensionless time $\tau/\pi$, evaluated with the Scenario-II mechanical core ($m_{\rm eff}=530$~ng, $f_m=100$~kHz) at illustrative couplings}. (a) Ideal vacuum QFI ($\alpha = 0$) vs cycles for different longitudinal Pauli couplings $k \in \{0.10, 0.20, 0.30\}$. (b) QFI vs cycles over the first mechanical cycle for different {initial coherent-state amplitudes} $|\alpha| \in \{1, 5, 10, 50\}$ at fixed coupling $k = 0.20$. (c) Linear entropy $S_L$ over the first mechanical cycle for different couplings $k \in \{0.10, 0.20, 0.30\}$ at qubit angle $\theta = \pi/2$.}
    \label{fig:qfi_linear_panel}
\end{figure*}

We consider an initial product state with the qubit prepared in a superposition $\cos(\theta/2)\ket{0} + \sin(\theta/2)\ket{1}$ and the mechanical resonator in a coherent state $\ket{\alpha}$ with mean phonon number $n = |\alpha|^2$.
\revred{The coherent amplitude $\alpha$ is defined in the fixed reference coordinate and is held locally independent of $g$ when taking the parameter derivative; this is a preparation assumption, separate from the coordinate shift.} A preparation that depends on $g$ requires the derivative of the initial mechanical state to be included in the QFI. The evolved state is Eq. \eqref{s9},
\begin{align}
|\Psi(\tau)\rangle
=&\, \cos\!\frac{\theta}{2}\, e^{i\phi_0(\tau)}
|0\rangle |\alpha_0(\tau)\rangle \notag\\
&+ \sin\!\frac{\theta}{2}\, e^{i\phi_1(\tau)}
|1\rangle |\alpha_1(\tau)\rangle,
\end{align}
where the conditional coherent amplitudes are \(|\alpha_j(\tau)\rangle = \big|\alpha e^{-i\tau} -\mathcal{Z}_j(1-e^{-i\tau})\big\rangle, \) with {$\mathcal{Z}_j = (-1)^{j+1} k + \bar G$}, and $\phi_j(\tau)$ represent the accumulated dynamical and geometric phases. The QFI sets the ultimate precision via the quantum Cram\'{e}r-Rao bound for the pure probe $\ket{\Psi(\tau)}$ as Eq.\eqref{s10}:
\begin{equation}
\mathcal{F}_Q(g)
= 4\!\left[
\langle \partial_g \Psi(\tau) | \partial_g \Psi(\tau) \rangle
- \big| \langle \Psi(\tau) | \partial_g \Psi(\tau) \rangle \big|^2
\right],
\label{eq:QFI_pure}
\end{equation}
With the dimensionless interaction time defined as $\tau=\omega_m t$ and the shorthand $\partial_g \equiv \tfrac{\partial}{\partial g}$, we consider a {superconducting gravimetric sensing model} with qubit populations $p_0=\cos^2(\theta/2)$ and $p_1=\sin^2(\theta/2)$, where the gravitational acceleration enters through the effective coupling $\bar G=\gamma g$, with $\gamma \equiv m_{\mathrm{eff}} z_{\mathrm{zpf}}/(\hbar\omega_m)$ and $z_{\mathrm{zpf}}=\sqrt{\hbar/(2m_{\mathrm{eff}}\omega_m)}$ setting the mechanical zero-point length scale. The longitudinal coupling is $k=g_z/\omega_m$, giving the qubit-conditioned mechanical shift {$\mathcal{Z}_j=(-1)^{j+1} k+\bar G$} for $j\in\{0,1\}$. The mechanical amplitudes then evolve as $\alpha_j(\tau)=\alpha e^{-i\tau}-\mathcal{Z}_j\eta(\tau)$ with $\eta(\tau)=1-e^{-i\tau}$, and for compactness we introduce {$A_j(\tau)=2\,\mathcal{Z}_j[\tau-\sin\tau] -\Im\!\big(\eta(\tau)\,\alpha^* e^{i\tau}\big)$}, allowing the gravitational QFI to be written in the closed form.
\begin{align}
\begin{split}
\mathcal{F}_Q(g;\tau)
= 4\gamma^2 \Bigg\{ &
\sum_{j=0}^1 p_j
\Big[ A_j^2 + |\eta|^2 ( 2|\alpha_j|^2 + 1 ) \\
& - 2(R_j^2-I_j^2) - 4 A_j I_j \Big] \\
& - \Bigg(
\sum_{j=0}^1 p_j \big( A_j - 2 I_j \big)
\Bigg)^2
\Bigg\},
\end{split}
\label{eq:QFI_closed}
\end{align}
where $R_j\equiv \Re(\eta\,\alpha_j^{*})$ and $I_j\equiv \Im(\eta\,\alpha_j^{*})$ capture the real and imaginary components of the qubit-conditioned phase-space displacement. {See Appendix~\ref{AA} for the detailed derivation.} \revred{Within the single-mode model of Eq.~\eqref{eq:H-longitudinal}, the mechanical trajectories recombine and the qubit and resonator momentarily disentangle at the stroboscopic revival times $\tau=2 n\pi$. For an equal superposition ($p_0=p_1=\tfrac12$), the QFI at these times reduces to}
\begin{equation}
\label{eq:FQ_revival}
\mathcal{F}_Q(g;\tau{=}2\pi n)
=256\,\pi^2 n^2\,\gamma^2\,k^2\,p_0 p_1
=64\,\pi^2 n^2\,\gamma^2\,k^2.
\end{equation}
\revred{The ideal reduced-qubit readout at an exact revival does not require the initial mechanical state to be locally independent of $g$.} \revred{At $T_n=2\pi n/\omega_m$, where $n=1,2,\ldots$ counts complete mechanical periods, the propagator in Eq.~\eqref{eq:evolution} factorizes as}
\begin{redrevision}
\begin{align}
\mathcal U(2\pi n)
&=e^{i\chi(g)}V_n(g)\otimes I_m,\nonumber\\
V_n(g)
&=\exp\!\left[-\frac{i}{2}
\bigl(\Omega_qT_n-8\pi nk\gamma g\bigr)\sigma_z\right],\nonumber\\
\chi(g)
&=2\pi n\bigl(k^2+\bar G^2\bigr).
\label{eq:revival-factorization}
\end{align}
\end{redrevision}
\revred{Here $I_m$ is the identity operator on the mechanical mode and $\chi(g)$ is an overall phase.}
\revred{For any initial product state $\rho_{q,0}\otimes\rho_m(g)$ with locally fixed qubit preparation, the reduced-qubit state is $\rho_q(T_n;g)=V_n(g)\rho_{q,0}V_n^\dagger(g)$.} \revred{For a balanced pure qubit superposition, the reduced-qubit QFI equals the value in Eq.~\eqref{eq:FQ_revival}, regardless of the mechanical input.} \revred{The mechanics returns to its initial state at the end of this ideal free interrogation. A gravity-dependent mechanical input retains information about $g$ in the joint state:}
\begin{redrevision}
\begin{equation}
F_Q^{\rm joint}(T_n)
=F_Q^{(q)}(T_n)+F_Q[\rho_m(g)].
\label{eq:revival-qfi-additivity}
\end{equation}
\end{redrevision}
\revred{Here $F_Q^{\rm joint}$ and $F_Q^{(q)}$ denote the joint and reduced-qubit QFIs, and $F_Q[\rho_m(g)]$ is the QFI of the initial mechanical state, all with respect to $g$.}

Eq.\eqref{eq:FQ_revival} makes explicit that the QFI grows quadratically with the longitudinal coupling strength $k$, and that maximal sensitivity is achieved when the qubit is prepared in a balanced superposition.
\revred{Away from the revival points, the full time-dependent QFI $\mathcal{F}_Q(g;\tau)$ exhibits oscillations at the mechanical frequency $\omega_m$.}
For any unbiased estimator $\hat{g}$ constructed from $N$ independent repetitions of the experiment~\cite{PhysRevLett.96.010401,Giovannetti2011}, the achievable precision is bounded by the quantum Cram\'{e}r--Rao inequality
\begin{equation}
\label{eq:QCRB_main}
\mathrm{Var}(\hat g)\;\ge\;\frac{1}{N\,\mathcal{F}_Q(g;\tau)},
\qquad
\Delta g_{\min}\;=\;\frac{1}{\sqrt{N\,\mathcal{F}_Q(g;\tau)}}.
\end{equation}
Assuming high-fidelity preparation, unitary evolution during the interrogation, and efficient projective readout, Eq.\eqref{eq:QCRB_main} depends only on the intrinsic probe dynamics and on the choice of interrogation time $\tau$. At the revival times $\tau=2 n \pi$, inserting Eq.\eqref{eq:FQ_revival} into Eq.\eqref{eq:QCRB_main} {gives the compact benchmark:}
\begin{equation}
\Delta g_{\min}(\tau{=}2 \pi n)
=\frac{1}{8\,\pi n\,|\gamma k|}\,\frac{1}{\sqrt{N}}
\label{eq:Delta_g_revival}
\end{equation}
We take the first-cycle peak value $\mathcal{F}_Q^{\mathrm{(peak)}}$ at time $t^\star$ and write:
\begin{equation}
\label{eq:Delta_g_from_peak}
\Delta g_{\min}(t^\star)
=\big[N\,\mathcal{F}_Q^{\mathrm{(peak)}}\big]^{-1/2}.
\end{equation}

Figure~\ref{fig:qfi_linear_panel} is {a closed-system diagnostic sweep, evaluated with the Scenario-II mechanical core ($m_{\rm eff}=530$~ng, $f_m=100$~kHz) at illustrative couplings}; the parameter-dependent simulations use Tables~\ref{tab:params-main} and~\ref{tab:scenarios_combined}. Fig.~\ref{fig:qfi_linear_panel} (a) shows the QFI evolution as a function of interaction time $\tau$ for various longitudinal coupling strengths $k$ (setting $\alpha = 0$). Here, the QFI quantifies the system's sensitivity to the gravitational acceleration $g$. As the system evolves, the mechanical trajectories associated with different qubit states separate in phase space.
\revred{The joint QFI measures distinguishability under changes in $g$, whereas the branch separation controls the overlap between the conditional mechanical states.} \revred{Since $\mathcal{Z}_1-\mathcal{Z}_0=2k$, the branch separation is independent of $g$ at fixed $k$ and $\tau$.} \revred{At fixed mechanical parameters and $\tau$, the magnitude of the derivative of the relative qubit phase with respect to $g$ scales with $|k|$.} \revred{Finally, the oscillations in the plot arise from the mechanical mode's periodicity, which also produces mechanical revivals at integer multiples of its period.}

Fig.~\ref{fig:qfi_linear_panel}(b) shows the QFI over the first mechanical cycle for several values of the common {initial coherent-state amplitude} $|\alpha|\in\{1,5,10,50\}$ at fixed $k=0.20$. The curves coincide: gravity couples linearly to the mechanical quadrature, so a uniform coherent displacement of both qubit-conditioned trajectories is a $g$-independent operation that leaves the QFI unchanged.
\revred{At exact revivals, the qubit QFI for fixed mechanical parameters is proportional to $k^2$.} \revred{Away from revival, the joint QFI also contains information from the common gravitational displacement of the oscillator.} Within this {closed-system comparison}, sensitivity increases with $k$; the circuit-constrained sweep in Fig.~\ref{fig:comparisonk} also changes $D_m$ and its associated flux noise.

Under the state-dependent displacement, the joint evolution entangles the qubit with the mechanical mode. Tracing out the mechanics yields a reduced qubit state whose coherence is set by the coherent-state overlap $O=\langle \alpha_1|\alpha_0\rangle$. The linear entropy is $\mathcal{S}_L = 1 - \mathrm{Tr}(\rho_q^2)$; for a qubit with excited-state population $p$ we obtain the closed form Eq.\eqref{s21} {(see Appendix~\ref{app:linearentropy})},
\begin{equation}
\mathcal{S}_L
\;=\; 2\,p(1-p)\,\bigl(1 - |O|^2\bigr).
\end{equation}
For conditional coherent states, we have $|O|^2 = \exp\!\bigl(-|\alpha_0 - \alpha_1|^2\bigr)$, where $\alpha_0 - \alpha_1 = (\mathcal{Z}_1 - \mathcal{Z}_0)\,(1 - e^{-i\tau})$. Thus, the entanglement is set by the phase-space separation controlled by the differential coupling. The periodicity follows from $|1 - e^{-i\tau}| = 2|\sin(\tau/2)|$, with $\mathcal{S}_L$ vanishing at $\tau = 2\pi n$ (full overlap) and reaching its maximum at $\tau = (2n+1)\pi$ (maximal separation). In the case of symmetric superpositions ($p = \tfrac{1}{2}$), $\mathcal{S}_L \to 1/2$ as $|O| \to 0$. Increasing $k$ increases the branch separation $|\alpha_0 - \alpha_1|$, reduces $|O|$, and raises the peak of $\mathcal{S}_L$ (see Fig. \ref{fig:qfi_linear_panel} c). In this sense, $\mathcal{S}_L$ provides a simple way to track how distinguishable the pointer states are during the cycle.

\revred{At $\tau=2\pi n$, the conditional mechanical trajectories close and the state factorizes. At these revivals, for the locally fixed mechanical input considered above, the joint-state QFI equals the reduced-qubit QFI carried by the gravity-dependent phase.} These revival points are therefore the stroboscopic qubit-readout points used for the reported benchmarks. Away from a revival, the joint-state QFI also includes information stored in the oscillator.

\subsection{Open-system dynamics}
\label{sec:decoherence}
In practice, hybrid qubit--mechanical devices are limited by environmental decoherence, which degrades both coherence and sensitivity. The system couples to its environment through several dissipation channels: the qubit undergoes energy relaxation and dephasing, and the mechanical resonator exchanges energy with its thermal bath. We describe the open-system dynamics with a Lindblad master equation for the laboratory-frame density matrix $\rho$; its polaron-frame form is given in Eq.~\eqref{s40}. \revred{Here, $\hat{\mathcal H}$ is the system Hamiltonian, and the dissipator $\mathcal{D}[\cdot]$ is defined in Eq.~\eqref{s41}.} The four collapse operators $\hat{L}_j$ describe the individual decoherence channels,
\begin{equation}
\begin{aligned}
    \hat{L}_1 &= \sqrt{\Gamma_1}\,\sigma_- ,
    &\qquad
    \hat{L}_2 &= \sqrt{\Gamma_\phi/2}\,\sigma_z ,\\
    \hat{L}_3 &= \sqrt{\gamma_m\!\left(n_{\mathrm{th}}+1\right)}\,\hat{a} ,
    &\qquad
    \hat{L}_4 &= \sqrt{\gamma_m\,n_{\mathrm{th}}}\,\hat{a}^\dagger .
\end{aligned}
\label{eq:collapse-ops}
\end{equation}
Here, $\Gamma_1$ is the qubit energy relaxation rate, often quoted via the relaxation time $T_1 = 1/\Gamma_1$. The physical intrinsic pure-dephasing rate is $\Gamma_\phi=1/T_\phi$; the factor $1/2$ in $L_2$ makes the off-diagonal density-matrix element decay at rate $\Gamma_\phi$. The mechanical damping rate is $\gamma_m$ (with quality factor $Q_m = \omega_m/\gamma_m$), and $n_{\mathrm{th}} = [\exp(\hbar\omega_m/k_B T) - 1]^{-1}$ is the thermal occupation of the mechanical bath at temperature $T$.

\revred{The stationary Gaussian treatment assumes a thermal mechanical state that is initially factorized from the qubit, centered at the fixed reference coordinate and locally independent of $g$. Under these assumptions and with Markovian qubit relaxation and background dephasing, the reduced laboratory-frame qubit state for an initial polar angle $\theta$ has the form:}
\begin{widetext}

\begin{equation}
\rho_q^{\mathrm{lab}}(t) =
\begin{pmatrix}
1 - \sin^2\!\tfrac{\theta}{2}\, e^{-\Gamma_1 t} &
\tfrac{1}{2}\sin\theta \; e^{+i\Phi(t)}\,e^{-\chi_m(t)}\,e^{-\Gamma_{2,{\rm bg}}t} \\[4pt]
\tfrac{1}{2}\sin\theta \; e^{-i\Phi(t)}\,e^{-\chi_m(t)}\,e^{-\Gamma_{2,{\rm bg}}t} &
\sin^2\!\tfrac{\theta}{2}\, e^{-\Gamma_1 t}
\end{pmatrix},
\label{eq:rho_lab}
\end{equation}

\end{widetext}
Here $\chi_m(t)$ is the stationary mechanical coherence cumulant defined in Eq.~\eqref{eq:mechanical-cumulant}. The background transverse-dephasing rate is
\begin{equation}
\Gamma_{2,{\rm bg}}
\equiv
\frac{\Gamma_1}{2}
+
\Gamma_\phi
\equiv
\frac{1}{T_{2R,{\rm bg}}}.
\label{eq:gamma2}
\end{equation}
The finite-time mechanical-bath contribution is carried by $\chi_m(t)$. The {dimensionless longitudinal coupling ratio} $k$ sets the strength of the qubit--mechanical interaction relative to the mechanical frequency.
\begin{figure*}[t!]
    \centering
    \begin{minipage}[t]{0.48\textwidth}
        \centering
        \textbf{Scenario I: Component-anchored Benchmark} \par\medskip
        \begin{subfigure}[b]{0.48\linewidth}
            \includegraphics[width=\linewidth]{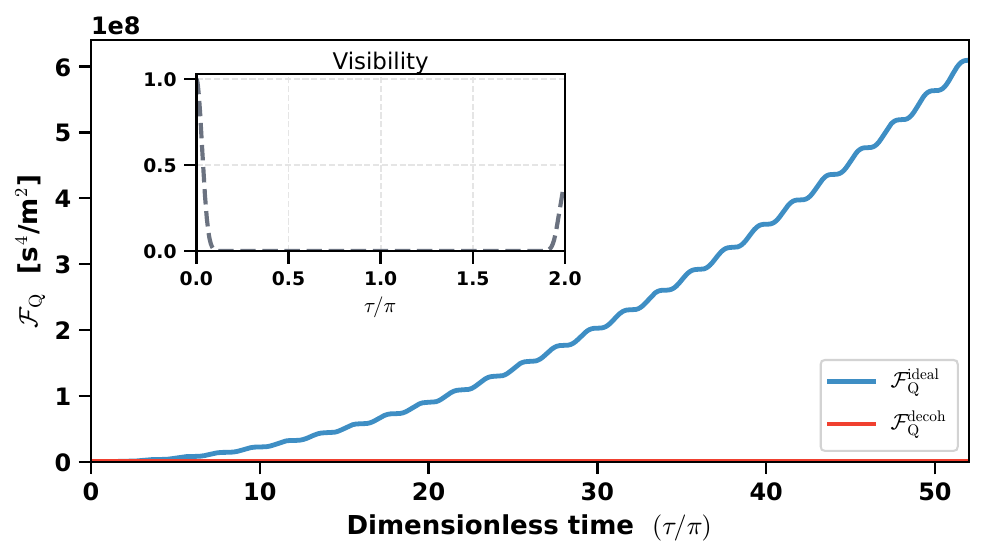}
            \caption{\label{fig:sc1-qfi}}
        \end{subfigure}\hfill
        \begin{subfigure}[b]{0.48\linewidth}
            \includegraphics[width=\linewidth]{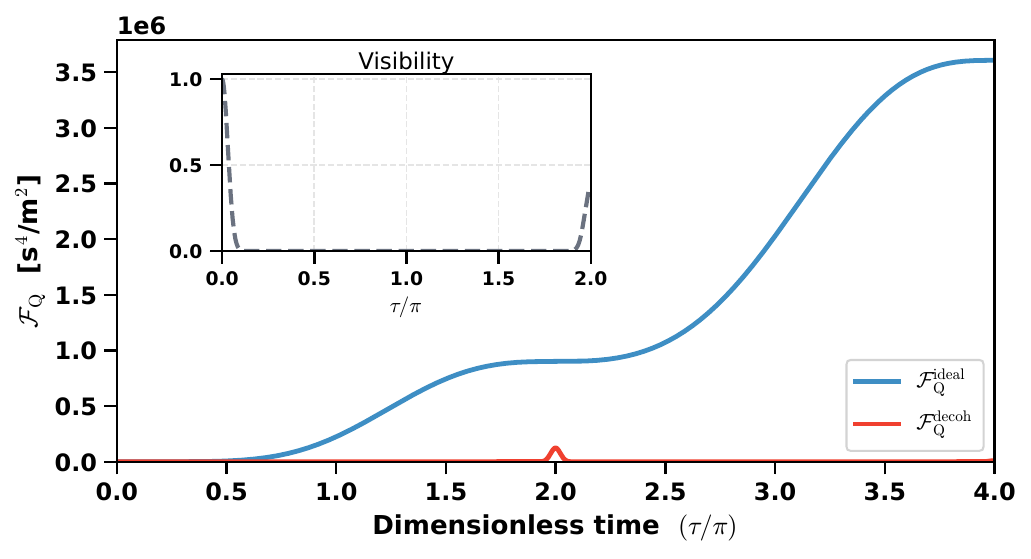}
            \caption{\label{fig:sc1-cycles}}
        \end{subfigure}
        \vspace{4pt}
        \begin{subfigure}[b]{0.98\linewidth}
            \includegraphics[width=\linewidth]{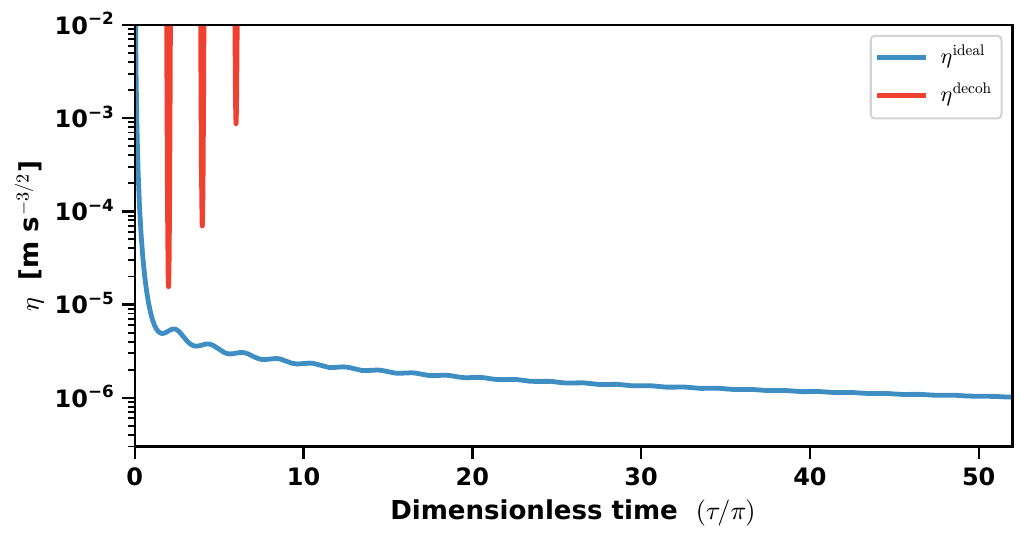}
            \caption{\label{fig:sc1-sens}}
        \end{subfigure}
    \end{minipage}
    \hfill \vrule \hfill
    \begin{minipage}[t]{0.48\textwidth}
        \centering
        \textbf{Scenario II: Long-term Integration Target} \par\medskip

        \begin{subfigure}[b]{0.48\linewidth}
            \includegraphics[width=\linewidth]{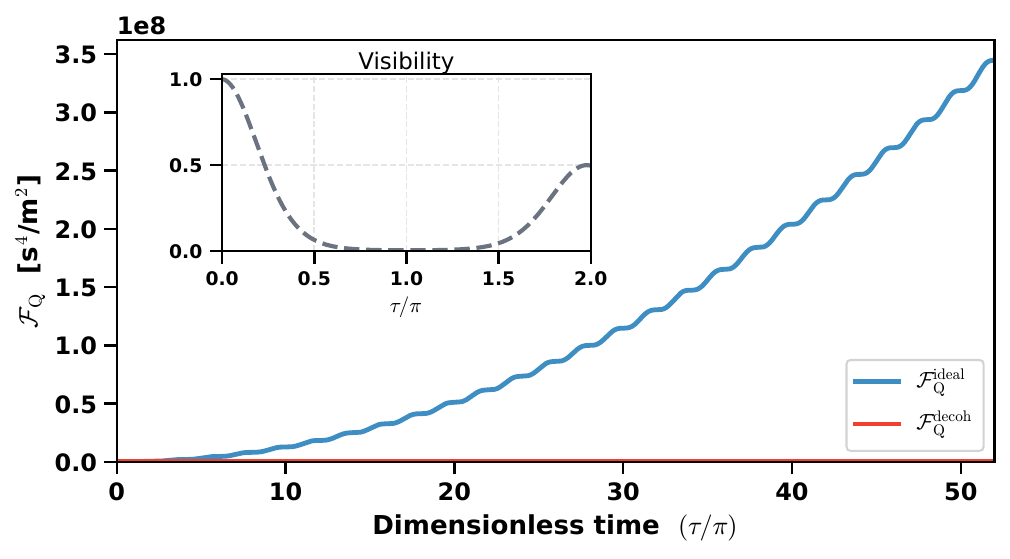}
            \caption{\label{fig:sc2-qfi}}
        \end{subfigure}\hfill
        \begin{subfigure}[b]{0.48\linewidth}
            \includegraphics[width=\linewidth]{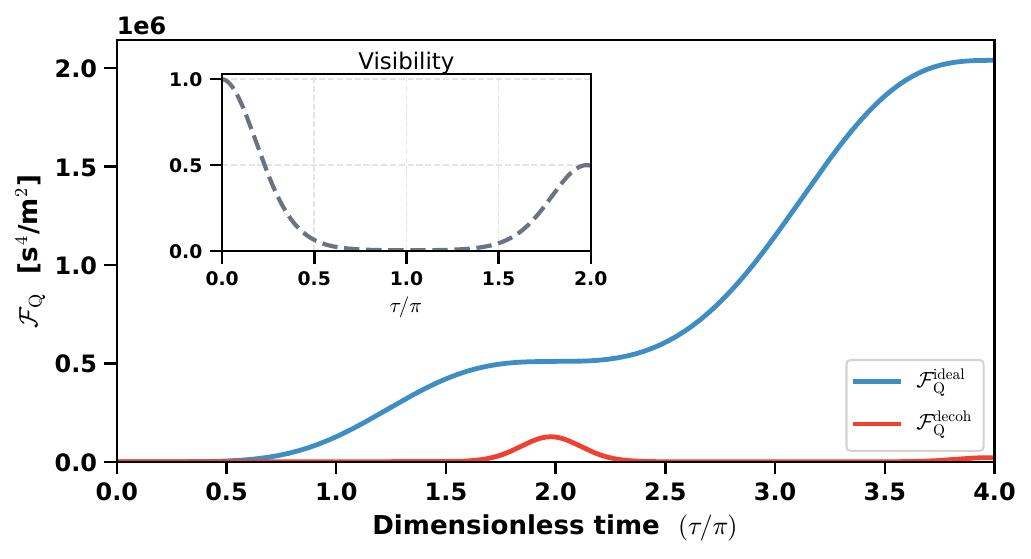}
            \caption{\label{fig:sc2-cycles}}
        \end{subfigure}

        \vspace{4pt}

        \begin{subfigure}[b]{0.98\linewidth}
            \includegraphics[width=\linewidth]{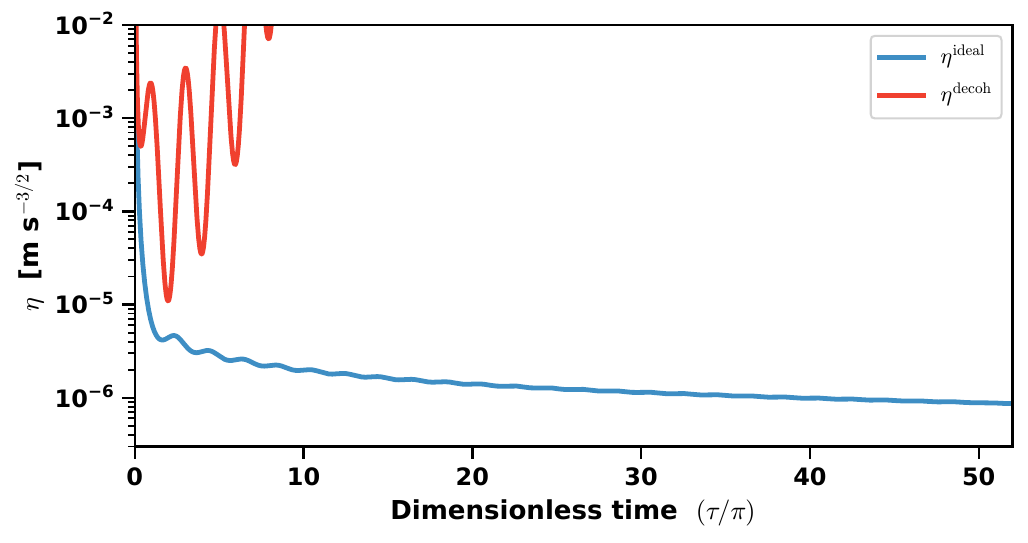}
            \caption{\label{fig:sc2-sens}}
        \end{subfigure}
    \end{minipage}
    \vspace{8pt}
    \caption{\justifying\textbf{Metrological performance comparison between device configurations.} \revred{The left column \textbf{(a)}--\textbf{(c)} displays Scenario I (component-anchored benchmark), while the right column \textbf{(d)}--\textbf{(f)} displays Scenario II (long-term integration target). All time-dependent curves use the locally fixed mechanical preparation assumed for the QFI in Eq.~\eqref{eq:FQ_full}.} \textbf{(a)}, \textbf{(d)} Evolution of the Quantum Fisher Information ($\mathcal{F}_Q$) over a large number of mechanical cycles. \textbf{(b)}, \textbf{(e)} Detailed view of $\mathcal{F}_Q$ in the short-time regime. \textbf{(c)}, \textbf{(f)} The corresponding independent-cycle, single-cycle-equivalent Fisher-rate sensitivity, $\eta_{g,{\rm ind}}$, as a function of interaction time. \revred{Blue curves show the $W(t)=1$ benchmark, which also omits mechanical which-path visibility loss away from ideal revivals.} {Red $\mathcal{F}_Q$ curves in (a), (b), (d), (e) include the coherence amplitude of Eq.~\eqref{eq:visibility-model}; the red sensitivity curves in (c), (f) additionally include readout contrast and cycle overhead. The insets show the full coherence amplitude $W(t)$. }}
    \label{fig:comparison}
\end{figure*}
For a stationary thermal mechanical mode with energy-damping rate $\gamma_m=\omega_m/Q_m$, define $s_m=\gamma_m/2+i\omega_m$. The Gaussian mechanical coherence cumulant is
\begin{equation}
\chi_m(t)=
4g_z^2(2n_{\rm th}+1)
\operatorname{Re}\!\left[
\frac{t}{s_m}
-\frac{1-e^{-s_m t}}{s_m^2}
\right].
\label{eq:mechanical-cumulant}
\end{equation}
In the lossless limit, Eq.~\eqref{eq:mechanical-cumulant} reduces to $4k^2(2n_{\rm th}+1)[1-\cos(\omega_m t)]$. The parameter-dependent figures use the coherence amplitude
\begin{equation}
W(t)=
\exp\!\left[
-\frac{t}{T_{2R,{\rm bg}}}
-\chi_\Phi(t)
-\chi_m(t)
\right].
\label{eq:visibility-model}
\end{equation}
Let $\delta\varphi(t)=\delta\Phi_m(t)/\Phi_0$. We adopt the two-sided flux-noise PSD convention
\begin{align}
\left\langle\delta\varphi(t)\delta\varphi(0)\right\rangle
={}&\int_{-\infty}^{\infty}df\,
S_\varphi^{(2)}(f)e^{i2\pi ft},\\
S_\varphi^{(2)}(f)={}&\frac{A_\varphi}{|f|},
\qquad f_{\rm IR}\leq |f|\leq f_{\rm UV},
\label{eq:flux-psd}
\end{align}
with $\sqrt{A_\varphi}=1.4\times10^{-6}$. Equivalently, the positive-frequency one-sided PSD is $S_\varphi^{(1)}(f)=2A_\varphi/f$. Consequently, $\sqrt{S_\varphi^{(1)}(1~\mathrm{Hz})}=\sqrt{2}\,\sqrt{A_\varphi}/\sqrt{\mathrm{Hz}}$; the quoted $0.9$, $1.2$, and $1.4$ values denote square roots of the spectral coefficient $A_\varphi$. The Gaussian Ramsey dephasing exponent is therefore
\begin{equation}
\chi_\Phi(t)=(2\pi D_m)^2A_\varphi
\int_{f_{\rm IR}}^{f_{\rm UV}}\frac{df}{f}
\left[\frac{\sin(\pi f t)}{\pi f}\right]^2.
\label{eq:flux-dephasing}
\end{equation}
Here $f_{\rm IR}=0.01~\mathrm{Hz}$, $f_{\rm UV}=1~\mathrm{MHz}$, and $T_{2R,{\rm bg}}=50~\mu\mathrm{s}$. The background Ramsey time excludes the explicitly modeled mechanical-loop $1/f$ flux noise and the stationary mechanical cumulant. The numerical curves use $\mathcal{F}_Q=W^2(\partial_g\Phi)^2$. The time-dependent phase $\Phi(t)$ carries the $g$--dependent signal. Separating phase and amplitude in this way makes it clear how signal build-up and decoherence compete over time. The precision limit for estimating $g$ is set by the QFI. Within the phase-covariant reduced-state model used here, $g$ enters through $\Phi(t)$, while $W(t)$, the qubit populations, and the decay rates are locally independent of $g$. The resulting $\mathcal{F}_Q$ is the qubit-accessible QFI and takes the form
\begin{equation}
\mathcal{F}_Q(g;t) = r_\perp^2(t)\,A^2(t),
\label{eq:QFI-simple}
\end{equation}
where $r_\perp(t)=\sin\theta\,W(t)$ for the numerical model is the transverse component of the Bloch vector, and $A(t) = \partial_g \Phi(t)$ is the phase sensitivity to $g$.
\revred{For the locally fixed mechanical input, substituting the high-$Q_m$ phase slope into Eq.~\eqref{eq:QFI-simple} gives the QFI used in the time-dependent numerical curves:}
\begin{widetext}
\begin{equation}
\mathcal{F}_Q(g;t) = \sin^2(\theta)\,
W^2(t)
\left[
\frac{4k}{\hbar \,\omega_m}\,
\sqrt{\frac{m_{\mathrm{eff}} \hbar}{2 \omega_m}}\,
\big(\sin(\omega_m t)-\omega_m t\big)
\right]^2.
\label{eq:FQ_full}
\end{equation}
\end{widetext}
At $\chi_\Phi(t)=0$, Eq.~\eqref{eq:FQ_full} reduces to the analytic open-system expression in Eq.~\eqref{s67}; Appendix~\ref{AD} gives the reduced-qubit derivation. To approach the QFI bound in practice, one must choose an appropriate measurement. For a Ramsey sequence with an adjustable local--oscillator phase $\varphi_{\mathrm{LO}}$, the classical Fisher information (CFI) quantifies how much information is extracted \cite{Helstrom1976QDET,Paris2009IJQI,Toth2014JPA,Pezze2018RMP}:
\begin{equation}
\mathcal{F}_C(g;t,\varphi_{\mathrm{LO}}) = \sum_{k=\pm} \frac{[\partial_g p_k(g)]^2}{p_k(g)},
\label{eq:CFI-def}
\end{equation}

\revred{In the absence of readout-assignment errors, the Ramsey outcome probabilities are}
\begin{redrevision}
\begin{equation}
p_\pm(g;t)=\frac12\left[1\pm r_\perp(t)
\cos\!\bigl(\Phi(g;t)+\varphi_{\rm LO}\bigr)\right].
\label{eq:ramsey-probabilities}
\end{equation}
\end{redrevision}
\revred{For the loaded input of Sec.~\ref{sec:loaded-preparation}, $\Phi$ is replaced by $\Phi_{\rm prep}$.} \revred{Recorded outcomes are fitted after applying the calibrated binary readout-assignment response to these probabilities.}

\revred{A baseline calibration or preliminary estimate $g_0$ sets $\varphi_{\rm LO}=\pi/2-\Phi(g_0;t)$, and this analysis-pulse phase is held fixed when taking $\partial_g$.} \revred{At $g=g_0$, the measurement is in quadrature and attains the local maximum CFI in Eq.~\eqref{s69}:}
\begin{equation}\label{eq19}
\mathcal{F}_C^{\max}(g;t)
=
r_\perp^2(t)\,A^2(t)
=
\mathcal{F}_Q(g;t),
\end{equation}
which saturates the quantum limit $\mathcal{F}_Q$ for an ideal Ramsey quadrature measurement in this phase-encoding model. The mechanical-bath exponent $\chi_m(t)$ scales with $g_z^2(2n_{\rm th}+1)$, so cryogenic operation remains essential. It is minimized near integer multiples of the mechanical period $2\pi/\omega_m$ and remains finite at an exact revival when $\gamma_m>0$. Finally, comparing the QFI in Eq.\eqref{eq:QFI-simple} with the CFI in Eq.\eqref{eq19} shows how finite purity degrades the achievable precision and gives a simple way to quantify the impact of decoherence in realistic settings.
\subsection{\revred{Preparation in a static gravitational field}}
\label{sec:loaded-preparation}

\revred{With gravity present, holding the qubit in $|0\rangle$ during the preparation described in Sec.~\ref{sec2} gives the stationary family of displaced thermal states under the local mechanical damping model,}
\begin{redrevision}
\begin{align}
\rho_m(g)&=D[\alpha_{\rm prep}(g)]\rho_{\rm th}
D^\dagger[\alpha_{\rm prep}(g)],\nonumber\\
\alpha_{\rm prep}(g)&=-\frac{\mathcal G-g_z}{\omega_m-i\gamma_m/2}
=\alpha_g+\alpha_{\rm bg},\nonumber\\
\alpha_g&=-\frac{\mathcal G}{\omega_m-i\gamma_m/2},
\qquad
\alpha_{\rm bg}=\frac{g_z}{\omega_m-i\gamma_m/2}.
\label{eq:loaded-preparation}
\end{align}
\end{redrevision}
\revred{Here $\rho_{\rm th}$ has the locally fixed covariance used in Eq.~\eqref{eq:mechanical-cumulant}, and $\alpha_{\rm bg}$ is locally independent of $g$.} \revred{An ideal short $\pi/2$ rotation prepares $|+\rangle\langle+|\otimes\rho_m(g)$, with $|+\rangle=(|0\rangle+|1\rangle)/\sqrt2$. This product form does not imply that the mechanical input is independent of gravity.}
\revred{With $s_m=\gamma_m/2+i\omega_m$, the common-force trajectory determining the coherent phase is}
\begin{redrevision}
\begin{equation}
\dot\alpha_c=-s_m\alpha_c-i\mathcal G,
\qquad
\alpha_c(t)=\alpha_g+\alpha_{\rm bg}e^{-s_mt}.
\label{eq:loaded-mean}
\end{equation}
\end{redrevision}
\revred{The corresponding Ramsey phase and its local derivative with respect to $g$ are}
\begin{redrevision}
\begin{align}
\Phi_{\rm prep}(t)&=\Omega_qt+
4g_z\operatorname{Re}\!\left[
\alpha_g t+\alpha_{\rm bg}\frac{1-e^{-s_mt}}{s_m}
\right],\nonumber\\
\partial_g\Phi_{\rm prep}(t)&=
-\frac{4\omega_m k\gamma\,t}{1+1/(4Q_m^2)}.
\label{eq:loaded-phase}
\end{align}
\end{redrevision}
\revred{The signal is the qubit frequency shift associated with the gravitational displacement already present before interrogation.}
\revred{A deterministic displacement leaves the connected thermal covariance unchanged. Within the same linear Gaussian model, with locally fixed bath covariance and noise parameters, $\chi_m(t)$ and $W(t)$ therefore remain those of Eqs.~\eqref{eq:mechanical-cumulant} and \eqref{eq:visibility-model}.} \revred{For the balanced qubit preparation, the reduced-qubit QFI is therefore}
\begin{redrevision}
\begin{equation}
F_Q^{(q)}(t)=W^2(t)\bigl[\partial_g\Phi_{\rm prep}(t)\bigr]^2.
\label{eq:loaded-qfi}
\end{equation}
\end{redrevision}
\revred{At $T_n=2\pi n/\omega_m$, the phase slope is $-8\pi nk\gamma/[1+1/(4Q_m^2)]$.} \revred{Relative to the high-$Q_m$ revival benchmark, the sensitivity is multiplied by $1+1/(4Q_m^2)$ under the same readout and cycle-time assumptions. The fractional correction is below $6\times10^{-13}$ for the listed quality factors, so the quoted revival values are unchanged to their stated precision.} \revred{The time-dependent figures use the locally fixed mechanical input assumed for the QFI in Eq.~\eqref{eq:FQ_full}.}

\subsection{\revred{Phase response and sensitivity scaling}}
\label{sec:addnotes}

We consider a qubit longitudinally coupled to a single mechanical mode of frequency $\omega_m$ with dimensionless coupling $k=g_z/\omega_m$. \revred{For the locally fixed mechanical input, the derivative of the high-$Q_m$ Ramsey phase with respect to $g$ is}
\begin{equation}
\partial_g\Phi(t)
=4k\Big(\frac{m_{\rm eff}z_{\rm zpf}}{\hbar\omega_m}\Big)
\big(\sin\tau-\tau\big),
\qquad \tau=\omega_m t.
\end{equation}
At short times, $\sin\tau-\tau=-\tau^3/6+O(\tau^5)$, so the phase sensitivity scales as $t^3$ and the corresponding qubit-accessible phase QFI scales as $F_Q^{(q)}\propto t^6$. \revred{For the loaded preparation of Sec.~\ref{sec:loaded-preparation}, the phase derivative with respect to $g$ is instead linear in time and gives $F_Q^{(q)}\propto t^2$ as $t\to0$.}

Mechanical which-path dephasing comes from the phase-space separation $\Delta\alpha(t)$ between the two qubit branches. For a thermal state with occupation $n_{\rm th}$, the thermal coherence amplitude and its squared visibility are $ W_{\rm th}(t)= \exp\!\left[-\frac{|\Delta\alpha(t)|^2}{2}(2n_{\rm th}+1)\right], \qquad \mathcal{D}_{\rm mech}(t)\equiv W_{\rm th}^2(t). $ Without damping, $\Delta\alpha(t)=2k\,(1-e^{-i\omega_m t})$, so $|\Delta\alpha|^2=16k^2\sin^2(\tau/2)$ and $ \mathcal{D}_{\rm mech}(t)=W_{\rm th}^2(t) =\exp[-16k^2(2n_{\rm th}+1)\sin^2(\tau/2)]. $ For a stationary damped thermal mode, the finite-temperature and damping contribution to the coherence amplitude is $e^{-\chi_m(t)}$. Background coherence, same-loop flux noise, and this mechanical factor are combined in $W(t)$ from Eq.~\eqref{eq:visibility-model}. The QFI for $g$ is then
\begin{equation}
F_Q(t)=\sin^2\theta\;W^2(t)\,\big[\partial_g\Phi(t)\big]^2.
\end{equation}
Because the lossless overlap term vanishes at $\tau=2\pi n$, and $\chi_m(t)$ is minimized nearby for high $Q_m$, useful interrogation times lie near $t^\star\simeq2\pi n/\omega_m$ whenever the total visibility remains appreciable.

\revred{We define $T_{\rm cyc}=t+T_{\rm over}$, where the total recurring overhead includes both Ramsey pulses, readout, qubit reset, intervening delays and any required mechanical restoration.} \revred{The assumed $T_{\rm over}=5\,\mu\mathrm{s}$ is a total recurring overhead budget requiring validation, not a passive mechanical equilibration time.}
\revred{At an ideal revival, readout is independent of the mechanical preparation and thus permits reuse of the resonator without complete re-equilibration before each shot.} \revred{Under noisy repetition, the Fisher-rate projections require the modeled Ramsey probabilities and their gravity dependence to remain valid despite changes in the mechanical input. Qubit reset alone does not establish this condition.} \revred{Assessing this repeatability condition requires accounting for mechanical evolution during all recurring operations. The present model does not establish a repeated-cycle implementation that meets both this condition and the assumed overhead budget.}
\revred{For symmetric binary readout-assignment errors with fidelity $F_r$, we use an effective Fisher information and its associated rate}
\begin{align}
F_{\rm eff}(t)&\approx(2F_r-1)^2F_Q(t),\\
R_F&\equiv\frac{F_{\rm eff}(t^\star)}{T_{\rm cyc}},\\
\eta_{g,{\rm ind}}&\equiv R_F^{-1/2}
=\sqrt{\frac{T_{\rm cyc}}{F_{\rm eff}(t^\star)}}.
\end{align}
\revred{$\eta_{g,{\rm ind}}$ is} the independent-cycle, single-cycle-equivalent Fisher-rate sensitivity. Equation~\eqref{eq:circuit-phase-slope} below and the numerical values of $\eta_{g,{\rm ind}}$ report the short-term Fisher-information rate under effectively independent repeated cycles. Low-frequency flux fluctuations can correlate successive cycles. Conversion to a measured one-second amplitude spectral density therefore requires the inter-cycle covariance, or equivalently an Allan-deviation characterization, of the integrated device. At the $n$th revival, substitution of Eq.~\eqref{eq:coupling-map} into the phase slope gives
\begin{redrevision}
\begin{equation}
\left|\partial_g\Phi(2\pi n/\omega_m)\right|
=8\pi n|k|\sqrt{\frac{m_{\rm eff}}{2\hbar\omega_m^3}}
=\frac{nD_m\lambda B\ell}{2\pi\Phi_0f_m^3}.
\label{eq:circuit-phase-slope}
\end{equation}
\end{redrevision}
{Substituting the circuit-derived coupling cancels the explicit $m_{\rm eff}$ factor from the ideal first-revival phase slope.} {An analogous mass cancellation was identified in nonlinear optomechanical gravimetry~\cite{Qvarfort2018}.} {The effective mass continues to enter the zero-point amplitude, $g_z$, and $\chi_m$, and it constrains the mechanical design.}

\section{Results and Discussion}
\label{sec:results}
{We evaluate the modeled sensing-head sensitivity using the QFI and the Cramer--Rao bound, including the specified decoherence channels, readout contrast, and cycle overhead.} Our analysis explores two distinct configurations. Scenario I is the component-anchored benchmark with $f_m=40.9\,\mathrm{kHz}$, $m_{\rm eff}=4.0\,\mathrm{ng}$, and $Q_m=6.6\times10^5$. Scenario II adopts a $100\,\mathrm{kHz}$, $530\,\mathrm{ng}$ mechanical core as a long-term integration target and uses $Q_m=10^6$. These parameters align with the Hamiltonian and open-system dynamics derived in Secs. \ref{sec2}--\ref{sec:decoherence}, with full control parameters and sensitivity metrics listed in Table \ref{tab:scenarios_combined}. Figure~\ref{fig:comparison} plots the time-dependent QFI and the independent-cycle, single-cycle-equivalent Fisher-rate sensitivity $\eta_{g,{\rm ind}}$ for both configurations. The values are evaluated at the first mechanical revival, where the closed-system conditional trajectories close and the stationary mechanical coherence loss is locally minimized.
\begin{table*}[t]
\caption{\justifying{\textbf{Model parameters and resulting performance metrics for Scenario~I and Scenario~II.}} \revred{Fixed parameters common to both scenarios are $T=20\,\mathrm{mK}$, $\sqrt{A_\varphi}=1.4\times10^{-6}$ under the two-sided PSD convention in Eq.~\eqref{eq:flux-psd}, $T_{2R,{\rm bg}}=50\,\mu\mathrm{s}$ and $F_r=0.995$, with the assumed total recurring overhead $T_{\rm over}=5\,\mu\mathrm{s}$ defined in Sec.~\ref{sec:addnotes}.} {The modeled-loss columns use the visibility in Eq.~\eqref{eq:visibility-model} and include readout contrast and cycle overhead.} {The ideal columns set the visibility and readout contrast to unity and omit overhead.}}
\label{tab:scenarios_combined}
\begin{ruledtabular}
\begin{tabular}{lcccc}
 & \multicolumn{2}{c}{\textbf{Scenario I}} & \multicolumn{2}{c}{\textbf{Scenario II}} \\
\textbf{Parameter / Metric} & {\textbf{Modeled losses}} & \textbf{Ideal} & {\textbf{Modeled losses}} & \textbf{Ideal} \\
 & \multicolumn{2}{c}{\textit{(component-anchored)}} & \multicolumn{2}{c}{\textit{(long-term target)}} \\
\hline
\multicolumn{5}{l}{\textit{Variable Parameters}} \\
Mechanical frequency $f_m$ (kHz) & 40.9 & 40.9 & 100 & 100 \\
Effective mass $m_{\mathrm{eff}}$ (ng) & 4.0 & 4.0 & 530 & 530 \\
Mechanical quality $Q_m$ & $6.6\times10^5$ & $6.6\times10^5$ & $10^6$ & $10^6$ \\
Field $B$ (mT) & 27 & 27 & 100 & 100 \\
Lever arm $\ell$ ($\mu$m) & 40 & 40 & 50 & 50 \\
Mechanical response $D_m$ (GHz) & 0.868 & 0.868 & 2.062 & 2.062 \\
Number coupling $g_n/2\pi$ (kHz) & 2.923 & 2.923 & 1.785 & 1.785 \\
Pauli coupling $g_z/2\pi$ (kHz) & 1.462 & 1.462 & 0.8926 & 0.8926 \\
Coupling ratio $k=g_z/\omega_m$ & 0.03574 & 0.03574 & 0.008926 & 0.008926 \\
\hline
\multicolumn{5}{l}{\textit{Performance Metrics}} \\
Selected revival $n$ & 1 & 1 & 1 & 1 \\
Interrogation time $t$ ($\mu$s) & 24.450 & 24.450 & 10.000 & 10.000 \\
Cycle time $T_{\rm cyc}$ ($\mu$s) & 29.450 & 24.450 & 15.000 & 10.000 \\
Visibility $W$ & 0.37195 & 1 & 0.49659 & 1 \\
Sensitivity $\eta_{g,{\rm ind}}$ ($\mathrm{m\,s^{-2}/\sqrt{Hz}}$) & $1.55\times10^{-5}$ & $5.21\times10^{-6}$ & $1.10\times10^{-5}$ & $4.43\times10^{-6}$ \\
State QFI $\mathcal{F}_Q(t)$ ($\mathrm{s^4/m^2}$) & $1.25\times10^5$ & $9.02\times10^5$ & $1.26\times10^5$ & $5.10\times10^5$ \\
\end{tabular}
\end{ruledtabular}
\end{table*}
The sensitivities in Table~\ref{tab:scenarios_combined} are conditional, independent-cycle, single-cycle-equivalent Fisher-rate projections for the {sensing head}. Scenario~I gives $\eta_{g,{\rm ind}}=1.55\times10^{-5}\,\mathrm{m\,s^{-2}/\sqrt{Hz}}$ at $t=24.450\,\mu\mathrm{s}$. Scenario~II gives $\eta_{g,{\rm ind}}=1.10\times10^{-5}\,\mathrm{m\,s^{-2}/\sqrt{Hz}}$ at $t=10.000\,\mu\mathrm{s}$. Joint integration at the listed $D_m$, field, mechanical, flux-noise, and coherence values remains to be demonstrated.

For comparison, define the single-mode quantum reference scale $a_{\rm SQL}=\omega_m\sqrt{\hbar/m_{\rm eff}}$. The Scenario-I and Scenario-II values in Table~\ref{tab:scenarios_combined} are, respectively, factors $11.8$ and $39.4$ larger than this scale. The circuit-constrained projections therefore lie above this single-mode quantum reference.

The oscillatory QFI structure in Fig. \ref{fig:comparison} directly reflects the stroboscopic protocol. The visibility revives at integer mechanical periods ($t = 2\pi n / \omega_m$), and the associated decoherence-limited QFI peaks lie close to those times. At the revival points the closed-system conditional trajectories close; finite mechanical damping leaves the residual coherence exponent in Eq.~\eqref{eq:mechanical-cumulant}. The red envelopes include background Ramsey decay, same-loop $1/f$ flux noise, and the stationary damped-mode mechanical cumulant. Their first-revival values produce the sensitivities quoted above.

Equation~\eqref{eq:circuit-phase-slope} fixes the interpretation of the long-term target: its projected value follows from the complete set $(f_m,B,\ell,D_m,Q_m)$ and the stated noise assumptions. The mass by itself supplies no independent gain in the circuit-constrained ideal phase slope. The figures therefore quantify {sensing-head} performance under these parameters; total gravimeter bandwidth and long-term resolution require measured instrument transfer functions.

\subsection{Design requirements for \texorpdfstring{$10^{-6}$}{10-6}-class internal sensitivity}
\label{sec:low-frequency-requirements}

At the nominal first mechanical revival, $t_1=1/f_m$, define the magnitude of the circuit-constrained phase slope by $s_g\equiv|\partial_g\Phi(t_1)|=qD_m$, where
\begin{equation}
q(f_m)=\frac{\lambda B\ell}{2\pi\Phi_0f_m^3}.
\label{eq:q-low-frequency}
\end{equation}
For fixed mechanical, coherence, and flux-noise parameters, the two $D_m$-dependent first-revival exponents in Eq.~\eqref{eq:visibility-model} have the form
\begin{equation}
\chi_\Phi(t_1)+\chi_m(t_1)=aD_m^2,
\qquad a>0,
\end{equation}
because both the flux-noise susceptibility and the Pauli coupling are linear in $D_m$. With $C_r=2F_r-1$ for the symmetric binary-assignment model, the single-cycle-equivalent Fisher-information-rate sensitivity for $D_m>0$ becomes
\begin{align}
\eta_{g,{\rm ind}}(D_m)
={}&\frac{\sqrt{T_{\rm cyc}}}{C_rqD_m}
\exp\!\left(\frac{1}{f_mT_{2R,{\rm bg}}}+aD_m^2\right).
\label{eq:eta-Dm-optimization}
\end{align}
Its unique minimum satisfies
\begin{align}
\frac{d\ln\eta_{g,{\rm ind}}}{dD_m}
={}&2aD_m-\frac{1}{D_m}=0,\notag\\
D_m^\star={}&\frac{1}{\sqrt{2a}},
\qquad a(D_m^\star)^2=\frac{1}{2}.
\label{eq:Dm-optimum}
\end{align}
Thus, within the fixed-parameter model, the useful response is set by the balance between phase gain and response-dependent decoherence. The optimum in Eq.~\eqref{eq:Dm-optimum} treats the response, background coherence, flux-noise amplitude, readout contrast, and circuit linearity as independent design coordinates; their bias-dependent correlations remain part of the integrated-device characterization.

\begin{figure*}[t]
\centering
\includegraphics[width=\textwidth]{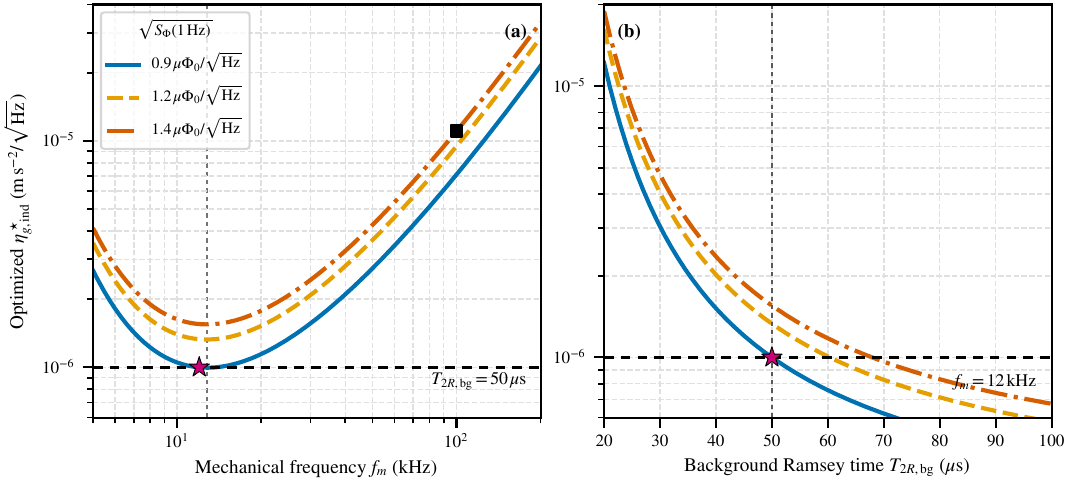}
\caption{\justifying\textbf{Model requirements for low-frequency operation.} At every point, $D_m$ is reoptimized using the stated two-sided PSD convention, the first-revival cycle model, and the stationary damped-mode cumulant in Eq.~\eqref{eq:mechanical-cumulant}. \textbf{(a)} Optimized single-cycle-equivalent sensitivity versus mechanical frequency at $T_{2R,{\rm bg}}=50~\mu\mathrm{s}$. The four curves use $\sqrt{A_\varphi}=(0.9,1.0,1.2,1.4)\times10^{-6}$. The $0.9$ value is the lower-end estimate, the $1.2$ and $1.4$ values are directly fitted coefficients from Ref.~\cite{Hutchings2017}, and $1.0\times10^{-6}$ is an intermediate model level. The gray vertical line marks the numerical minimum at $12.80~\mathrm{kHz}$ for the $\sqrt{A_\varphi}=0.9\times10^{-6}$ curve. The black square marks the reoptimized Scenario-II point at $100~\mathrm{kHz}$ and $\sqrt{A_\varphi}=1.4\times10^{-6}$; the magenta star marks the rounded requirements point in Table~\ref{tab:low-frequency-target}. \textbf{(b)} Optimized single-cycle-equivalent sensitivity versus background Ramsey time at $f_m=12~\mathrm{kHz}$; the gray vertical line marks $T_{2R,{\rm bg}}=50~\mu\mathrm{s}$. The horizontal dashed line in each panel marks $10^{-6}~\mathrm{m\,s^{-2}/\sqrt{Hz}}$.}
\label{fig:low-frequency-requirements}
\end{figure*}

In the flux-dominated regime, the Ramsey filter integral gives $a\propto A_\varphi f_m^{-2}$ up to its logarithmic cutoff factor. Upon neglecting cycle overhead in the leading power law, Eq.~\eqref{eq:eta-Dm-optimization} therefore yields
\begin{equation}
\eta_{g,{\rm ind}}^\star\propto
\frac{\sqrt{A_\varphi}\,f_m^{3/2}}{\lambda B\ell}
\exp\!\left(\frac{1}{f_mT_{2R,{\rm bg}}}\right).
\label{eq:optimized-frequency-scaling}
\end{equation}
The corresponding analytic minimum is $f_m^{\rm opt}\simeq2/(3T_{2R,{\rm bg}})=13.3~\mathrm{kHz}$ for $T_{2R,{\rm bg}}=50~\mu\mathrm{s}$. The numerical model, including the stated flux-noise cutoffs, the $5~\mu\mathrm{s}$ cycle overhead, and the stationary damped-mode cumulant, gives $f_m^{\rm opt}=12.80~\mathrm{kHz}$ {[Fig.~\ref{fig:low-frequency-requirements}(a)]}.

{Table~\ref{tab:low-frequency-target} specifies a rounded requirements point at $f_m=12~\mathrm{kHz}$.} It inherits the Scenario-II field, lever arm, mass, quality factor, temperature, readout, and overhead; adopts the lower-end normalized spectral coefficient $\sqrt{A_\varphi}=0.9\times10^{-6}$; and reoptimizes $D_m$. The model evaluation gives
\begin{equation}
\eta_{g,{\rm ind}}=1.00\times10^{-6}\,
\mathrm{m\,s^{-2}/\sqrt{Hz}},
\label{eq:eta-low-frequency-target}
\end{equation}
rounded to the precision of the requirements inputs. {This value is a conditional projection of the sensing head's internal sensitivity, based on the single-cycle-equivalent Fisher-information rate for statistically independent cycles.} Conversion to a measured one-second amplitude spectral density requires the inter-cycle covariance, active tracking of low-frequency offsets, or an Allan-deviation characterization of the integrated device.

Here ``internal'' includes the modeled qubit--mechanical signal, residual background Ramsey decay, mechanical-loop flux noise, thermal mechanical coherence, mechanical damping, binary-readout contrast, and cycle overhead. Instrument-level performance additionally requires measured transfer functions for vibration, tilt, temperature-to-flux drift, magnetic-field drift, absolute calibration, multimode response, and long-time stability.

\revred{The modeled readout cannot distinguish gravity from platform acceleration producing the same mechanical loading. Mechanical isolation can reduce vibration, and synchronized auxiliary motion measurements can support correction through calibrated transfer functions. Leveling and independent orientation measurements can support tilt correction. Rotation relative to the magnetic field can also alter loop flux and responsivity, so orientation calibration must include the chip--field geometry.}

The Ramsey integral uses $f_{\rm IR}=0.01~\mathrm{Hz}$ and $f_{\rm UV}=1~\mathrm{MHz}$. These cutoffs and the independent-cycle rate conversion are separate assumptions of the reported metric.

\begin{table}[t]
\caption{\justifying\textbf{Requirements point for $10^{-6}$-class internal sensitivity.} Fixed values inherited from Scenario~II are $m_{\rm eff}=530~\mathrm{ng}$, $Q_m=10^6$, $B=100~\mathrm{mT}$, $\ell=50~\mu\mathrm{m}$, $\lambda=0.9$, $T=20~\mathrm{mK}$, $F_r=0.995$, and $T_{\rm over}=5~\mu\mathrm{s}$. The optimized response and derived quantities use Eqs.~\eqref{eq:mechanical-cumulant} and~\eqref{eq:Dm-optimum}.}
\label{tab:low-frequency-target}
\centering
\footnotesize
\setlength{\tabcolsep}{4pt}
\begin{ruledtabular}
\begin{tabular}{lc}
Quantity & Requirements value \\
\hline
Mechanical frequency $f_m$ & $12.0~\mathrm{kHz}$ \\
Flux-noise amplitude $\sqrt{A_\varphi}$ & $0.9\times10^{-6}$ \\
Background coherence $T_{2R,{\rm bg}}$ & $50~\mu\mathrm{s}$ \\
Optimized response $D_m^\star$ & $0.4135~\mathrm{GHz}$ \\
Number coupling $g_n/2\pi$ & $1.034~\mathrm{kHz}$ \\
Pauli coupling $g_z/2\pi$ & $0.5168~\mathrm{kHz}$ \\
Coupling ratio $k=g_z/\omega_m$ & $0.04307$ \\
Interrogation time $t_1$ & $83.333~\mu\mathrm{s}$ \\
Cycle time $T_{\rm cyc}$ & $88.333~\mu\mathrm{s}$ \\
Phase slope $|\partial_g\Phi(t_1)|$ & $8.288\times10^4~\mathrm{s^2/m}$ \\
Flux-noise exponent $\chi_\Phi(t_1)$ & $0.49676$ \\
Mechanical exponent $\chi_m(t_1)$ & $0.003238$ \\
Visibility $W(t_1)$ & $0.11456$ \\
Sensitivity $\eta_{g,{\rm ind}}$
& $1.00\times10^{-6}~\mathrm{m\,s^{-2}/\sqrt{Hz}}$ \\
\end{tabular}
\end{ruledtabular}
\end{table}

The experimental support for this point is component-level and spans distinct device architectures. Room-temperature optomechanical trampolines have been measured over $10$--$200~\mathrm{kHz}$ with masses of order $100~\mathrm{ng}$ and quality factors reaching $9\times10^5$~\cite{Kleckner2011}. \revred{The SQUID experiments summarized in Table~\ref{tab:mechanical-evidence} support $Q_m=10^6$ as a plausible mechanical design target~\cite{Rodrigues2019Flux,Bera2024}.} \revred{Achieving this value at $100~\mathrm{mT}$ after superconducting metallization and SQUID integration remains unverified for the proposed $100$- and $12$-kHz designs, both with $m_{\rm eff}=530~\mathrm{ng}$.}

\begin{table*}[t]
\begingroup
\color{RevisionRed}
\caption{\justifying\textbf{Reported mechanical parameters and adopted model values.} The experimental fields are in-plane fields. Literature quality factors are inferred from resonance-linewidth measurements under the listed conditions. The sensing model uses the energy-damping quality factor $Q_m=\omega_m/\gamma_m$. The complete model inputs and conditional sensitivity projections are given in Tables~\ref{tab:scenarios_combined} and~\ref{tab:low-frequency-target}.}
\label{tab:mechanical-evidence}
\centering
\small
\setlength{\tabcolsep}{4pt}
\renewcommand{\arraystretch}{1.12}
\begin{ruledtabular}
\begin{tabular}{@{}
>{\raggedright\arraybackslash}p{0.20\textwidth}
>{\centering\arraybackslash}p{0.12\textwidth}
>{\centering\arraybackslash}p{0.08\textwidth}
>{\centering\arraybackslash}p{0.16\textwidth}
>{\raggedright\arraybackslash}p{\dimexpr0.44\textwidth-8\tabcolsep\relax}
@{}}
Source and status
& \shortstack[c]{Frequency $f_m$\\(kHz)}
& \shortstack[c]{Field $B$\\(mT)}
& Quality factor
& Device and operating conditions \\
\hline
\multicolumn{5}{@{}l}{\textit{Reported integrated-device measurements}} \\
\hline
Rodrigues \emph{et al.}~\cite{Rodrigues2019Flux}
& $7129$
& $9$
& $Q_m\simeq9\times10^5$
& Aluminum beam forming a SQUID arm in a microwave circuit. Reported intrinsic mechanical linewidth near $8~\mathrm{Hz}$. \\
Bera \emph{et al.}~\cite{Bera2024}
& $3.97\times10^3$
& $\simeq18$
& $Q_m\simeq6.6\times10^5$
& Aluminum mechanical arm in a transmon SQUID. Intrinsic linewidth near $6~\mathrm{Hz}$ inferred under weak pumping and reduced polariton flux responsivity. \\
Poot \emph{et al.}~\cite{Poot2010}
& $1053$
& $100$
& $Q_0\simeq5300$
& Mechanical resonator integrated into a voltage-state niobium dc-SQUID with proximity junctions. Quoted $Q_0$ is intrinsic. Operating mechanical damping depends on detector bias. \\
\hline
\multicolumn{5}{@{}l}{\textit{Adopted model inputs and mechanical requirements}} \\
\hline
Scenario~I: composite benchmark
& $40.9$
& $27$
& $Q_m=6.6\times10^5$
& $m_{\rm eff}=4~\mathrm{ng}$. Combines mechanical and SQUID results from separate devices and operating conditions, as described in Sec.~\ref{sec2}. \\
Scenario~II: requirements
& $100$
& $100$
& $Q_m=10^6$
& $m_{\rm eff}=530~\mathrm{ng}$. Mechanical requirement for operation at $100~\mathrm{mT}$ after superconducting metallization and SQUID integration. \\
Lower-frequency requirements point
& $12$
& $100$
& $Q_m=10^6$
& $m_{\rm eff}=530~\mathrm{ng}$. Inherits the Scenario-II mass, field and quality-factor requirements for the lower-frequency design. \\
\end{tabular}
\end{ruledtabular}
\endgroup
\end{table*}

Hutchings \emph{et al.} obtained $\sqrt{A_\Phi}=1.2$--$1.4~\mu\Phi_0$ from direct fits and estimated $0.9~\mu\Phi_0$ from the scatter of their weakly tunable sample \cite{Hutchings2017}. These values are the coefficients $\sqrt{A_\Phi}$ in the two-sided model; the corresponding one-sided one-hertz amplitude spectral densities are larger by $\sqrt{2}$. Their Ramsey extraction used a $1~\mathrm{Hz}$ infrared cutoff. The present combination of the estimate $\sqrt{A_\Phi}=0.9~\mu\Phi_0$ with $f_{\rm IR}=0.01~\mathrm{Hz}$ is a requirements assumption. Reoptimizing $D_m$ with the directly fitted $\sqrt{A_\Phi}=1.2$ and $1.4~\mu\Phi_0$ coefficients gives $1.33\times10^{-6}$ and $1.55\times10^{-6}~\mathrm{m\,s^{-2}/\sqrt{Hz}}$, respectively.

The target response $D_m^\star=0.4135~\mathrm{GHz}$ corresponds to $|\partial f_q/\partial\Phi_m|=0.4135~\mathrm{GHz}/\Phi_0$, within the sub-$1~\mathrm{GHz}/\Phi_0$ response scale used by Hutchings \emph{et al.} to suppress flux-dependent dephasing. Its realization for the mechanical-loop coordinate at the stated field and coherence remains an integrated-circuit requirement. \revred{Poot \emph{et al.} demonstrated integrated mechanical transduction at $100~\mathrm{mT}$ with a $50$-$\mu$m arm and a $1.053$-MHz resonator. The resonator was read out by a voltage-state niobium dc-SQUID with proximity junctions~\cite{Poot2010}.} \revred{They reported intrinsic $Q_0\simeq5300$ and bias-dependent mechanical damping.} The value $\lambda=0.9$ serves as a geometry benchmark for the fundamental mode of that transducer. For the fabricated target, $\lambda$, $m_{\rm eff}$, and $z_{\rm zpf}$ must be calculated using the same normalized mechanical generalized coordinate; the resulting mode shape then determines the target coupling. Thin-film aluminum transmons retain microsecond coherence to at least $0.65~\mathrm{T}$~\cite{Krause2022}. A recent gradiometric-transmon preprint reports $T_2^*\simeq33~\mu\mathrm{s}$ at the sweet spot and values above $30~\mu\mathrm{s}$ under small flux bias \cite{Fu2026}. Together these results provide separate field-resilience and coherence precedents; $T_{2R,{\rm bg}}=50~\mu\mathrm{s}$ at $100~\mathrm{mT}$ in the mechanically integrated circuit is part of the requirements vector {[Fig.~\ref{fig:low-frequency-requirements}(b)]}.

\revred{Untracked perpendicular-field drift can produce a Ramsey phase that mimics acceleration. Trapped vortices can increase circuit loss and flux noise and degrade qubit coherence or readout. Vortex trapping depends on device geometry and field history, including the cooling procedure.}

The target number-coupling ratio $g_n/\omega_m=(g_n/2\pi)/f_m=0.0861$ is about $2.2$ times the estimated maximum dressed-polariton ratio near $4\%$ reported by Bera \emph{et al.}, and about $7.6$--$8.6$ times their directly fitted ratios \cite{Bera2024}. The target absolute number coupling is $g_n/2\pi=1.03~\mathrm{kHz}$. This cross-architecture comparison is an order-of-magnitude integration benchmark. Joint realization of this normalized coupling, the mechanical target, the field, the stated flux noise, the background coherence, and mechanical thermalization to $20~\mathrm{mK}$ defines the integrated-device requirements vector.

\revred{At fixed transmon-loop bias, define the reference curvature $c_m=\partial_\varphi^2 f_q$ and the ratio $\mathcal R_m=|c_m|/D_m$, where $\varphi=\Phi_m/\Phi_0$.} \revred{When quadratic curvature is the leading correction, limiting the variation in responsivity and longitudinal coupling to $1\%$ requires $|c_m\Delta\varphi|/D_m\lesssim0.01$.} \revred{Once the mechanics has settled under the residual loading, the harmonic high-$Q_m$ model gives}
\begin{redrevision}
\begin{align}
|\Delta\varphi|
&\simeq
\frac{\lambda|B|\ell}{\Phi_0\omega_m^2}|g|,
\nonumber\\
|g|
&\lesssim
\frac{0.01}{\mathcal R_m}
\frac{\Phi_0\omega_m^2}{\lambda|B|\ell}.
\label{eq:residual-linearity-range}
\end{align}
\end{redrevision}
\revred{For the parameters in Tables~\ref{tab:scenarios_combined} and~\ref{tab:low-frequency-target}, this gives}
\begin{redrevision}
\begin{equation}
|g|\lesssim
\begin{cases}
1.405/\mathcal R_{\mathrm I},
& \text{Scenario I},\\
1.814/\mathcal R_{\mathrm{II}},
& \text{Scenario II},\\
0.02612/\mathcal R_{12},
& f_m=12~\mathrm{kHz},
\end{cases}
\quad \mathrm{m\,s^{-2}}.
\label{eq:residual-linearity-examples}
\end{equation}
\end{redrevision}
\revred{A definite numerical interval requires $c_m$ at each operating point and cannot be determined from $D_m$ alone.} \revred{The relevant thermal and qubit-conditioned mechanical excursions must also remain within the linear-response interval, with higher-order corrections negligible.} \revred{Fixed-phase Ramsey readout is additionally restricted by the fringe in Eq.~\eqref{eq:ramsey-probabilities}. The quoted sensitivities assume local quadrature operation.}

The readout value $F_r=0.995$ has a separate fixed-frequency-transmon component precedent at a $140~\mathrm{ns}$ readout time \cite{Chen2023Readout}. \revred{The assumed $5~\mu\mathrm{s}$ total recurring overhead budget is defined in Sec.~\ref{sec:addnotes} and remains an integration requirement.} Sunada \emph{et al.} demonstrated a $40~\mathrm{ns}$ readout probe pulse and a $100~\mathrm{ns}$ reset as separate operations~\cite{Sunada2022}. Their integration with the sensing interaction and the symmetric binary-assignment model forms part of the readout requirement.

At the target, $n_{\rm th}=3.47\times10^4$. For a small timing offset $\delta\tau=\omega_m\delta t$ about the first revival, the lossless thermal overlap has the quadratic form
\begin{equation}
W_{\rm th}\simeq
\exp[-2k^2(2n_{\rm th}+1)\delta\tau^2].
\end{equation}
The condition $W_{\rm th}\geq0.99$ gives $|\delta t|\leq82.8~\mathrm{ns}$. This corresponds to $|\delta f_m|\leq11.9~\mathrm{Hz}$ at $12~\mathrm{kHz}$, provided the frequency mismatch is approximately constant during each interrogation and pulse-timing error is negligible. \revred{Fixed timing is sufficient when the combined phase mismatch from frequency fluctuations and pulse-timing errors satisfies this thermal-overlap criterion. Otherwise, period tracking, stabilization, or improved pulse timing is needed.} For the same point, the one-sided classical viscous-damping convention gives
\begin{equation}
S_{a,{\rm th}}^{1/2}
=\sqrt{\frac{4k_BT\omega_m}{m_{\rm eff}Q_m}}
=1.25\times10^{-8}~\mathrm{m\,s^{-2}/\sqrt{Hz}}.
\end{equation}
Structural damping produces a frequency-dependent input-noise spectrum \cite{Saulson1990}. \revred{The single-mode reference is $a_{\rm SQL}=\omega_m\sqrt{\hbar/m_{\rm eff}} =3.36\times10^{-8}~\mathrm{m\,s^{-2}/\sqrt{Hz}}$.} The projected $\eta_{g,{\rm ind}}$ lies above these two reference scales by factors $79.8$ and $29.7$, respectively.

\revred{At the first nominal revival $t_1=1/f_m$, the high-$Q_m$ limit of Eq.~\eqref{eq:mechanical-cumulant}, with $\gamma_m t_1\ll1$, gives}
\begin{redrevision}
\[
\chi_m(t_1)\simeq
\frac{8\pi k^2(2n_{\rm th}+1)}{Q_m}.
\]
\end{redrevision}
\revred{At the requirements point in Table~\ref{tab:low-frequency-target}, this exponent is $3.2\times10^{-3}$.} \revred{Reducing $Q_m$ from $10^6$ to $10^5$, while keeping $D_m$ and the other model inputs at their tabulated values, increases $\eta_{g,{\rm ind}}$ to $1.03\times10^{-6}~\mathrm{m\,s^{-2}/\sqrt{Hz}}$.} \revred{This comparison quantifies an isolated damping change. It does not estimate the degradation caused by SQUID integration, superconducting metallization, or operation at $100~\mathrm{mT}$.}

\revred{We vary the stationary mechanical temperature $T$, evaluating the initial connected thermal covariance and bath occupation at the same temperature.} \revred{At $f_m=12~\mathrm{kHz}$ and $Q_m=10^6$, we reoptimize $D_m$ at each temperature using Eq.~\eqref{eq:Dm-optimum}, with the remaining inputs in Table~\ref{tab:low-frequency-target} held fixed. The resulting sensitivities $\eta_{g,{\rm ind}}$ are $(1.00,\,1.01,\,1.03)\times10^{-6}~\mathrm{m\,s^{-2}/\sqrt{Hz}}$ at $T=20$, $70$, and $200~\mathrm{mK}$, respectively.} \revred{At the reoptimized $70~\mathrm{mK}$ point, the lossless thermal-overlap condition $W_{\rm th}\geq0.99$ gives $|\delta t|\leq44.6~\mathrm{ns}$.} \revred{This comparison excludes nonthermal fluctuations and temperature-induced changes in damping, circuit coherence, flux noise, and readout.}

\revred{In a separate variation, we reduce $T_{2R,{\rm bg}}$ to $30~\mu\mathrm{s}$ while keeping $Q_m=10^6$ and the remaining inputs at their values for the requirements point. At $f_m=12~\mathrm{kHz}$, this gives $\eta_{g,{\rm ind}}=3.0\times10^{-6}~\mathrm{m\,s^{-2}/\sqrt{Hz}}$.} \revred{Jointly reoptimizing $f_m$ and $D_m$ for this coherence time, with the other inputs unchanged, gives $\eta_{g,{\rm ind}}=2.2\times10^{-6}~ \mathrm{m\,s^{-2}/\sqrt{Hz}}$ at $f_m\simeq21~\mathrm{kHz}$.}

\revred{Table~\ref{tab:gravimeter-comparison} places the conditional sensing-head projections alongside representative reported gravimeter benchmarks, with the measurement conditions and temporal characteristics identified for each instrument.} \revred{The nominal cycle rates for the present projections do not establish tracking bandwidth.}

\begin{table*}[t]
\begingroup
\color{RevisionRed}
\caption{Representative gravimeter benchmarks and conditional sensing-head Fisher-rate projections with modeled losses. Noise and sensitivity values retain the source definitions. Spectral self-noise at a specified frequency and short-term averaging coefficients are different metrics. Manufacturer specifications are identified separately from research measurements. Recording rates, cycle rates, interrogation times and measured bandwidths are distinguished.}
\label{tab:gravimeter-comparison}
\centering
\small
\setlength{\tabcolsep}{4pt}
\renewcommand{\arraystretch}{1.12}
\begin{ruledtabular}
\begin{tabular}{@{}
>{\raggedright\arraybackslash}p{0.17\textwidth}
>{\raggedright\arraybackslash}p{0.23\textwidth}
>{\raggedright\arraybackslash}p{0.27\textwidth}
>{\raggedright\arraybackslash}p{\dimexpr0.33\textwidth-6\tabcolsep\relax}
@{}}
\shortstack[l]{Platform and\\source}
& \shortstack[l]{Noise or sensitivity\\($\mathrm{m\,s^{-2}/\sqrt{Hz}}$)}
& Timing, rate or bandwidth
& Operating conditions and status \\
\hline
\multicolumn{4}{@{}l}{\textit{Reported gravimeter benchmarks}} \\
\hline
Superconducting iGrav~\cite{GWRiGravSpec}
& $3.0\times10^{-9}$\newline
Specified noise. Frequency band not stated.
& Continuous relative recording.\newline
Bandwidth not specified in the cited specification.
& $4.2~\mathrm{K}$ helium cryostat.\newline
Manufacturer specification. \\
\hline
MOEMS, Wu \emph{et al.}~\cite{Wu2025}
& $1.1\times10^{-8}$\newline
Self-noise at $0.5~\mathrm{Hz}$.
& Sampling rate for noise-spectrum measurement: $10~\mathrm{Hz}$.\newline
Tide recording: $0.1~\mathrm{Hz}$.
& Temperature-controlled cave laboratory. Optical readout without vibration isolation or vacuum maintenance. Measured. \\
\hline
Force-balanced MEMS, Gao \emph{et al.}~\cite{Gao2026}
& $1.0\times10^{-9}$\newline
Self-noise at $0.14~\mathrm{Hz}$.
& Measured closed-loop bandwidth: $\mathrm{DC}$--$108~\mathrm{Hz}$.\newline
Initial tide-recording rate: $1~\mathrm{Hz}$.
& Vacuum, temperature and tilt control. Self-noise extracted using coherence analysis to remove microtremor signals. Measured. \\
\hline
Cold-atom GAIN, Freier \emph{et al.}~\cite{Freier2016GAIN}
& $9.6\times10^{-8}$\newline
Reported short-term noise.
& Raman-pulse separation: $T=260~\mathrm{ms}$.\newline
Total interval: $2T=520~\mathrm{ms}$.\newline
Cycle: $1.5~\mathrm{s}$.
& $^{87}\mathrm{Rb}$ atoms at $2~\mu\mathrm{K}$. Vacuum and active vibration isolation. Measured. \\
\hline
Cold-atom AQG, M\'enoret \emph{et al.}~\cite{Menoret2018AQG}
& $5.0\times10^{-7}$\newline
Best reported short-term sensitivity.
& Raman-pulse separation: $T=60~\mathrm{ms}$.\newline
Total interval: $2T=120~\mathrm{ms}$.\newline
Cycle rate: approximately $2~\mathrm{Hz}$.
& Instrument range $18$--$30~{}^\circ\mathrm{C}$. Atoms below $2~\mu\mathrm{K}$ in vacuum, with vibration compensation. Measured. \\
\hline
Free-fall FG5-X~\cite{FG5XSpec,Freier2016GAIN}
& $1.5\times10^{-7}$\newline
Specified precision coefficient at a quiet site.
& Drop rate: $0.1~\mathrm{Hz}$ in the comparison campaign of Ref.~\cite{Freier2016GAIN}.
& $20$--$30~{}^\circ\mathrm{C}$. Vacuum and active isolation. Manufacturer specification. Recording rate from the campaign. \\
\hline
\multicolumn{4}{@{}l}{\textit{Conditional sensing-head projections}\textsuperscript{a}} \\
\hline
Present Scenario~I
& $1.55\times10^{-5}$\newline
Conditional $\eta_{g,{\rm ind}}$.
& Interrogation: $24.450~\mu\mathrm{s}$.\newline
Cycle: $29.450~\mu\mathrm{s}$.\newline
Nominal cycle rate: $33.96~\mathrm{kHz}$.
& Assumed mechanical temperature $20~\mathrm{mK}$.\newline
Component-based projection. \\
\hline
Present Scenario~II
& $1.10\times10^{-5}$\newline
Conditional $\eta_{g,{\rm ind}}$.
& Interrogation: $10.000~\mu\mathrm{s}$.\newline
Cycle: $15.000~\mu\mathrm{s}$.\newline
Nominal cycle rate: $66.67~\mathrm{kHz}$.
& Assumed mechanical temperature $20~\mathrm{mK}$.\newline
Prospective integrated-device projection. \\
\hline
Present requirements point
& $1.00\times10^{-6}$\newline
Conditional $\eta_{g,{\rm ind}}$.
& Interrogation: $83.333~\mu\mathrm{s}$.\newline
Cycle: $88.333~\mu\mathrm{s}$.\newline
Nominal cycle rate: $11.32~\mathrm{kHz}$.
& Assumed mechanical temperature $20~\mathrm{mK}$.\newline
Projection assuming the joint design requirements are met. \\
\end{tabular}
\end{ruledtabular}
\smallskip
\begin{minipage}{\textwidth}
\footnotesize\raggedright
\textsuperscript{a)} The present $\eta_{g,{\rm ind}}$ projections assume repeatable input statistics, a full recurring overhead of $5~\mu\mathrm{s}$, and effectively independent cycles, as in Tables~\ref{tab:scenarios_combined} and~\ref{tab:low-frequency-target}. Their nominal cycle rates do not establish tracking bandwidth.
\end{minipage}
\endgroup
\end{table*}

\section{Conclusion}
\label{conclusion}

{We model a proposed chip-integrated superconducting gravimetric sensing head comprising a flux-tunable transmon longitudinally coupled to a co-fabricated nanomechanical beam through a flux-mediated interaction. From the effective longitudinal Hamiltonian, we obtain an exact factorized propagator for the closed system and identify the acceleration-dependent qubit phase. We derive the qubit-accessible QFI for the unitary dynamics and the reduced open-system model. At mechanical revivals, the closed-system conditional trajectories close in phase space, allowing the acceleration-dependent phase to be read out with partially recovered Ramsey visibility.}

{The transmon-frequency responsivity to mechanical-loop flux controls the acceleration-dependent phase signal and responsivity-dependent decoherence. With all other model parameters fixed, the acceleration-to-phase slope at the first revival scales linearly with this responsivity. The associated flux-noise and mechanical-bath dephasing exponents scale quadratically. This signal--decoherence tradeoff produces a finite optimal flux responsivity. The optimum sets a measurable target for device design.}

{At the first mechanical revival, the component-anchored benchmark gives a projected sensitivity of $1.6\times10^{-5}~\mathrm{m\,s^{-2}/\sqrt{Hz}}$ at $24.5~\mu\mathrm{s}$; the prospective integrated-device parameter set gives $1.1\times10^{-5}~\mathrm{m\,s^{-2}/\sqrt{Hz}}$ at $10~\mu\mathrm{s}$.} {Using the lower-end flux-noise coefficient, a separate requirements analysis identifies a rounded $12~\mathrm{kHz}$ requirements point. At this requirements point, the optimized responsivity is $D_m^\star=0.414~\mathrm{GHz}$, and the projected sensitivity is $1.0\times10^{-6}~\mathrm{m\,s^{-2}/\sqrt{Hz}}$. With the higher flux-noise coefficients obtained from direct fits, the projected sensitivity lies in the range $(1.3\text{--}1.6)\times10^{-6}~ \mathrm{m\,s^{-2}/\sqrt{Hz}}$. The joint requirements for an integrated device encompass the mechanical parameters, transmon flux responsivity, magnetic field, flux-noise amplitude, coherence properties, cycle timing, and readout fidelity. These sensitivity projections are conditional on the specified device and noise parameters. They use the single-cycle-equivalent Fisher-information rate and assume statistically independent cycles.}

{The mechanical-bath contribution to qubit-coherence loss depends on the mechanical quality factor and bath temperature and helps constrain the useful interrogation window. Within the phase-encoding model, an ideal Ramsey quadrature measurement saturates the qubit-accessible QFI. The sensitivity projections incorporate qubit relaxation and dephasing, mechanical damping, same-loop low-frequency flux-noise dephasing, binary-readout contrast, and cycle overhead.} {The analysis applies to a chip-integrated sensing head operated within a dilution refrigerator. Establishing instrument-level vibration rejection, thermal stability, absolute calibration, bandwidth, and long-term stability requires experimental characterization of the full cryogenic instrument.}

\section*{Acknowledgement}
This publication was partially supported by the Qatar Research, Development and Innovation (QRDI) Council under the Academic Research Grant ARG01-0603-230468. The findings and views expressed herein are solely the responsibility of the authors.
\section*{Declarations}

\subsection*{Ethics approval and consent to participate}
Not applicable. This study is purely theoretical and computational; it did not involve human participants, human data, or animals, and therefore did not require approval from an ethics committee or institutional review board.

\subsection*{Availability of data and materials}
The data and simulation code that support the findings of this study are available from the corresponding author upon reasonable request.

\subsection*{Competing interests}
The authors declare that they have no competing interests.

\bibliographystyle{apsrev4-2}
\bibliography{references}
\begin{widetext}
\newpage
\appendix
\label{app}
\section{Quantum Fisher Information for Closed System}
\label{AA}
In this Appendix we first derive the exact unitary dynamics and the corresponding Quantum Fisher Information (QFI) for the closed qubit--oscillator system. The starting point is the Hamiltonian
\begin{align}\label{s1}
\begin{split}
    \mathcal{H} =& \frac{\hbar\Omega_q}{2}\sigma_{z}
+ \hbar\omega_m\,a^{\dagger}a \\&
+ \hbar g_z\,\sigma_{z}(a+a^{\dagger})
+ m_{\mathrm{eff}}g\,z_{\mathrm{zpf}}\,(a+a^{\dagger})\,.
\end{split}
\end{align}
Here, $\Omega_q$ is the qubit frequency, $\omega_m$ is the mechanical frequency, $g_z$ is the Pauli longitudinal coupling strength between the qubit and the mechanical mode, and $m_{\mathrm{eff}}g\,z_{\mathrm{zpf}}$ represents the static gravitational force projected onto the mechanical zero-point motion. The parameter $z_{\mathrm{zpf}} = \sqrt{\hbar/(2m_{\mathrm{eff}}\omega_m)}$ is the zero-point fluctuation amplitude of the mechanical mode. For convenience, we now introduce dimensionless parameters that collect the relevant energy scales and define a dimensionless time:
\begin{equation}\label{s2}
    k = \frac{g_z}{\omega_m},\quad
\bar{G} = \frac{m_{\mathrm{eff}}g z_{\mathrm{zpf}}}{\hbar\omega_m},\quad
\tau = \omega_m t,\quad
\mathcal{Z} = k\sigma_z + \bar{G}.
\end{equation}
In terms of these quantities the Hamiltonian can be written as
\begin{equation}\label{s3}
    \mathcal{H} = \hbar\omega_m\bigl[a^\dagger a+\mathcal{Z}(a+a^\dagger)\bigr]
+\tfrac{\hbar\Omega_q}{2}\sigma_z.
\end{equation}
This form makes explicit that the mechanical oscillator is displaced by the operator $\mathcal{Z}$, which contains both the qubit-dependent longitudinal coupling and the gravity-induced contribution. To diagonalize the Hamiltonian in the mechanical sector, we perform a polaron (displacement) transformation. The corresponding unitary is defined as
\begin{equation}
\mathcal{T} = \exp\left[-\mathcal{Z}(a^{\dagger} - a) \right].
\end{equation}
Under this transformation, the mechanical ladder operators are shifted according to
\begin{align}\label{s5}
\mathcal{T}^{\dagger} a \mathcal{T} &= a - \mathcal{Z}, &
\mathcal{T}^{\dagger} a^{\dagger} \mathcal{T} &= a^{\dagger} - \mathcal{Z}.
\end{align}
Applying the transformation $\mathcal{H}' = \mathcal{T}^{\dagger} \mathcal{H} \mathcal{T}$ yields an effective Hamiltonian in which the longitudinal coupling and gravitational force are absorbed into a renormalized energy shift:
\begin{align}
\mathcal{H}' = \hbar\omega_m \left[ a^{\dagger}a - \mathcal{Z}^2 \right] + \frac{\hbar\Omega_q}{2}\sigma_z.
\end{align}
The terms in $\mathcal{H}'$ mutually commute, since $[a^\dagger a, \sigma_z] = [a^\dagger a, \mathcal{Z}^2] = [\sigma_z, \mathcal{Z}^2] = 0$. As a result, the time-evolution operator in the polaron frame factorizes as
\begin{align}
    \begin{split}
        \mathcal{U}'(t) =& \exp(-i \tau a^\dagger a) \exp(i \tau \mathcal{Z}^2)\\& \exp\left(-i \frac{\Omega_q}{2\omega_m} \tau \sigma_z\right).
    \end{split}
\end{align}
To return to the original (laboratory) frame, we conjugate by $\mathcal{T}$ and simplify the resulting expression. Using the Baker--Campbell--Hausdorff (BCH) formula for the conjugation of free evolution by a displacement operator, the full evolution operator takes the standard form
\begin{align}\label{s4}
\begin{split}
\mathcal{U}(t)
=&\exp\Bigl(-i\frac{\Omega_q}{2 \omega_m} \tau \sigma_z\Bigr)
\exp\Bigl[i\mathcal{Z}^2(\tau-\sin\tau)\Bigr]
\\& \times \exp\Bigl[-\mathcal{Z}\bigl((1-e^{-i\tau})a^\dagger-(1-e^{i\tau})a\bigr)\Bigr]\\& \times
\exp(-i\tau\,a^\dagger a).
\end{split}
\end{align}
The structure of Eq.\eqref{s4} clearly separates the free qubit precession, the gravity- and coupling-dependent phase factor, the qubit- and gravity-conditioned displacement of the mechanical oscillator, and the free mechanical evolution. We now apply this unitary evolution to a factorized initial state, where the qubit is prepared in a superposition state and the resonator in a coherent state:
\begin{equation}
    \ket{\Psi(0)}=(\cos\tfrac\theta2\ket0+\sin\tfrac\theta2\ket1)\ket\alpha.
\end{equation}
Using Eq.\eqref{s4}, the evolved state can be written as
\begin{align}\label{s9}
    \begin{split}
    \ket{\Psi(\tau)} =& \cos\frac\theta2\,e^{i\phi_0(\tau)}\ket0\ket{\alpha_0(\tau)} \\&+\sin\frac\theta2\,e^{i\phi_1(\tau)}\ket1 \ket{\alpha_1(\tau)},
    \end{split}
\end{align}
which shows explicitly that the qubit populations label two distinct mechanical trajectories in phase space. Here,
\begin{equation}
\begin{aligned}
\alpha_j(\tau) &= \alpha e^{-i \tau} - \mathcal{Z}_j \left(1 - e^{-i \tau}\right), \\
\phi_j(\tau) &= (-1)^j \frac{\Omega_q}{2 \omega_m} \tau + \mathcal{Z}_j^2 \left(\tau - \sin(\tau)\right) + \chi_j(\tau).
\end{aligned}
\end{equation}
with {$\mathcal{Z}_j = (-1)^{j+1} k + \bar{G}$}. The additional phase $\chi_j(\tau)$ arises from the coherent-state displacement and is given by
\begin{equation}
\begin{aligned}
\chi_j(\tau) &= {\frac{1}{2i}} \bigl[\beta_j (\alpha e^{-i\tau})^* - \beta_j^* (\alpha e^{-i\tau})\bigr], \\
\beta_j &= -\mathcal{Z}_j \left(1 - e^{-i\tau}\right), \qquad
\alpha_j(\tau) = \alpha e^{-i \tau} + \beta_j.
\end{aligned}
\end{equation}
Equations above fully specify the ideal (closed-system) qubit--oscillator state as a function of the gravitational acceleration $g$. For a pure state $\ket{\Psi(\tau)}$ that depends on the parameter $g$, the QFI is given by the standard expression
\begin{equation}\label{s10}
    \mathcal{I}_g =4\Bigl[\langle\partial_g\Psi|\partial_g\Psi\rangle-|\langle\Psi|\partial_g\Psi\rangle|^2\Bigr],
\end{equation}
which quantifies the ultimate precision bound for unbiased estimators of $g$ based on this state. To obtain an explicit closed form, we introduce the shorthand \( \gamma={m_{\rm eff}z_{\rm zpf}}/{\hbar\omega_m},\quad \eta(\tau)=1-e^{-i\tau}, \) and write the derivative of the state with respect to $g$ as \( \partial_g\ket{\Psi} =\gamma\sum_{j=0,1}c_j e^{i\phi_j(\tau)} \ket{j}\otimes\ket{\dot\alpha_j}, \) where $c_0=\cos(\theta/2)$, $c_1=\sin(\theta/2)$, and
\[
\ket{\dot\alpha_j}
=\bigl(iA_j+{\Re(\eta\,\alpha_j^*)}\bigr)\ket{\alpha_j}
-\eta\,a^\dagger\ket{\alpha_j},
\quad
A_j=\partial_{\mathcal{Z}_j}\phi_j.
\]
Substituting these expressions into Eq.\eqref{s10} and performing the coherent-state algebra yields the QFI in the compact form
\begin{align}\label{s11}
    \begin{split}
        \mathcal{F}_Q=& 4\gamma^2\Biggl\{\sum_j|c_j|^2\Bigl[A_j^2+|\eta|^2(2|\alpha_j|^2+1)
    \\&-2\Re[(\eta\alpha_j^*)^2] -4A_j\Im(\eta\alpha_j^*)\Bigr]
    \\&-\Bigl[\sum_j|c_j|^2(A_j-2\Im(\eta\alpha_j^*))\Bigr]^2\Biggr\}.
    \end{split}
\end{align}
Equation~\eqref{s11} is the QFI of the joint qubit--mechanical pure state. Away from a revival it includes information stored in the oscillator and can exceed the QFI accessible through qubit-only Ramsey readout. \revred{At the closed-system revivals $\tau=2\pi n$, the conditional mechanical trajectories close and the state factorizes. For the locally fixed mechanical input used here, the joint-state and reduced-qubit QFIs coincide at these revivals.} The open-system performance results in the main text use the reduced-qubit QFI.
\section{Calculation of Linear Entropy}
\label{app:linearentropy}
We next quantify the entanglement generated between the qubit and the mechanical oscillator by evaluating the linear entropy of the reduced qubit state. Consider a general qubit--oscillator pure state of the form
\begin{equation}
|\Psi(t)\rangle = \sum_{j \in \{0,1\}} c_j(t) |j\rangle \otimes |\alpha_j(t)\rangle,
\end{equation}
where $|j\rangle$ denotes the computational basis of the qubit and $|\alpha_j(t)\rangle$ are coherent states of the oscillator with complex amplitudes $\alpha_j(t)$. The corresponding total density matrix is
\begin{equation}
\rho(t) = \sum_{j,k \in \{0,1\}} c_j(t) c_k^*(t)\, |j\rangle\langle k| \otimes |\alpha_j(t)\rangle\langle\alpha_k(t)|.
\end{equation}
Tracing over the oscillator degrees of freedom yields the reduced qubit state:
\begin{equation}
\rho_q(t) = \mathrm{Tr}_{\text{osc}}[\rho(t)]
= \sum_{j,k \in \{0,1\}} c_j(t) c_k^*(t)\,
\langle\alpha_k(t)|\alpha_j(t)\rangle\, |j\rangle\langle k|.
\end{equation}
The key quantity controlling decoherence in this subsystem is the overlap between the two coherent states $\ket{\alpha_0(t)}$ and $\ket{\alpha_1(t)}$. Introducing the overlap integral
\begin{align}
    \begin{split}
        \mathcal{O}(t) = \exp \bigg(&-\frac{1}{2}\left|\alpha_1(t) - \alpha_0(t)\right|^2 \\&
        + \frac{1}{2}\left[\alpha_1^*(t)\alpha_0(t)
        - \alpha_0^*(t)\alpha_1(t)\right]\bigg),
    \end{split}
\end{align}
and using the standard identities for coherent states, its magnitude is:
\begin{equation}\label{s16}
\left|\mathcal{O}(t)\right| = \exp\left(-\frac{1}{2}\left|\alpha_1(t) - \alpha_0(t)\right|^2\right).
\end{equation}
In the $\{|0\rangle,|1\rangle\}$ basis, the reduced density matrix can therefore be written as
\begin{equation}
\rho_q(t) = \begin{pmatrix}
|c_0(t)|^2 & c_0(t) c_1^*(t)\, \mathcal{O}(t) \\
c_1(t) c_0^*(t)\, \mathcal{O}^*(t) & |c_1(t)|^2
\end{pmatrix}.
\end{equation}
The purity of the reduced state, $\mathrm{Tr}[\rho_q^2(t)]$, quantifies how mixed the qubit has become due to entanglement with the oscillator:
\begin{align}
    \begin{split}
        \mathrm{Tr}\!\left[\rho_q^2(t)\right]
        =& |c_0(t)|^4 + |c_1(t)|^4 \\&
        + 2|c_0(t)|^2 |c_1(t)|^2 \left|\mathcal{O}(t)\right|^2.
    \end{split}
\end{align}
The linear entropy,
\begin{equation}
    \mathcal{S}_L(t) = 1 - \mathrm{Tr}\!\left[\rho_q^2(t)\right],
\end{equation}
then provides a simple entanglement monotone for the bipartite pure state. Substituting the purity expression yields
\begin{align}
    \begin{split}
        \mathcal{S}_L(t) = 1 - \Big(&|c_0(t)|^4 + |c_1(t)|^4 \\&
        + 2|c_0(t)|^2 |c_1(t)|^2 \left|\mathcal{O}(t)\right|^2\Big).
    \end{split}
\end{align}
Introducing the population parameter $p \equiv |c_0(t)|^2$ (so that $|c_1(t)|^2 = 1-p$ by normalization), we obtain the compact form
\begin{equation}\label{s21}
    \mathcal{S}_L(t) = 2p(1-p)\left(1 - \left|\mathcal{O}(t)\right|^2\right).
\end{equation}
The linear entropy thus increases as the coherent-state overlap $|\mathcal{O}(t)|^2$ decreases and vanishes when the two conditional trajectories fully overlap. To connect this entanglement measure to the gravitationally induced dynamics of our model, we now specify the coherent-state amplitudes for a representative longitudinally coupled evolution. In the main text and in Sec.~\ref{AA}, the qubit eigenstates $|j\rangle$ are associated with dimensionless displacement parameters $\mathcal{Z}_j$ through the operator $\mathcal{Z} = k\sigma_z + \bar{G}$. For a simple revival model we take
\begin{equation}\label{s22}
\alpha_j(t) = {-}\mathcal{Z}_j\!\left(1 - e^{-i\omega_m t}\right),
\end{equation}
where $\mathcal{Z}_j$ is the state-dependent displacement amplitude for $|j\rangle$ and $\omega_m$ is the mechanical frequency. The amplitude difference is therefore
\begin{equation}
\alpha_0(t) - \alpha_1(t) = {\left(\mathcal{Z}_1 - \mathcal{Z}_0\right)}\!\left(1 - e^{-i\omega_m t}\right).
\end{equation}
Substituting this into the expression for the linear entropy gives
\begin{align}
    \begin{split}
        \mathcal{S}_L(t) =& 2p(1-p) \\& \times \left[1 - \exp\!\left(-\left|\mathcal{Z}_0 - \mathcal{Z}_1\right|^2\left|1 - e^{-i\omega_m t}\right|^2\right)\right].
    \end{split}
\end{align}
At the first mechanical revival, $t=2\pi/\omega_m$ and $\tau=\omega_m t=2\pi$. Hence $1-e^{-i\tau}=0$, $|\mathcal{O}(t)|^2=1$, and $\mathcal{S}_L(t)=0$. The vanishing of the linear entropy at integer multiples of the period $2\pi/\omega_m$ indicates complete disentanglement between qubit and oscillator, demonstrating periodic revivals of separability in the joint dynamics.

\revred{Since $\mathcal{Z}_1-\mathcal{Z}_0=2k$, the overlap magnitude and $\mathcal{S}_L$ are independent of $g$ at fixed $k$, $\tau$ and $p$.} \revred{The phase factor in Eq.~\eqref{s4} contributes a gravity-dependent relative qubit phase through $\mathcal{Z}_1^2-\mathcal{Z}_0^2=4k\bar G$.} \revred{This phase remains available for Ramsey readout when the conditional trajectories close at exact revivals in the ideal harmonic model.}
\section{Lindblad Derivation}
We now extend the analysis to an open-system setting by deriving an effective Lindblad master equation for the qubit in the presence of mechanical dissipation and qubit decoherence. The system Hamiltonian in the laboratory frame is given by Eq.~\eqref{s1}. As in the closed-system case, we implement the polaron transformation via the unitary operator
\begin{equation}
\mathcal{T} = \exp[-\mathcal{Z}({a}^\dagger - {a})],
\end{equation}
where the displacement parameter $\mathcal{Z}$ is defined in Eq.\eqref{s2}. Under this transformation, the mechanical operators become
\begin{equation}
\mathcal{T}^\dagger{a}\mathcal{T} = {a} - \mathcal{Z}, \qquad \mathcal{T}^\dagger{a}^\dagger\mathcal{T} = {a}^\dagger - \mathcal{Z},
\end{equation}
The longitudinal operator is unchanged, $\mathcal{T}^\dagger\sigma_z\mathcal{T}=\sigma_z$. The ladder operators acquire mechanical displacements,
\begin{equation}
\mathcal{T}^\dagger\sigma_\pm\mathcal{T}
=\sigma_\pm\exp[\pm2k(a^\dagger-a)].
\end{equation}
Consequently, $\sigma_x$ and $\sigma_y$ are dressed by mechanical displacement operators. Applying the polaron transformation to the full Hamiltonian leads to
\begin{equation}
\mathcal{H}' = \mathcal{T}^\dagger\mathcal{H}\mathcal{T} = \frac{\hbar}{2}\tilde{\Omega}_q\sigma_z + \hbar\omega_m{a}^\dagger{a},
\end{equation}
where the qubit frequency is renormalized by the longitudinal coupling and gravity as \( \tilde{\Omega}_q = \Omega_q - 4\omega_m k\bar{G} \). A constant energy shift $-\hbar\omega_m(k^2 + \bar{G}^2)$ has been omitted, as it does not affect the dynamics. The system couples to multiple dissipative channels described by Lindblad operators. In the laboratory frame these are taken as:
\begin{equation}
\begin{aligned}
 L_1 &= \sqrt{\Gamma_1}\,\sigma_- \otimes \mathbb{I}_m,
&
 L_2 &= \sqrt{\Gamma_\phi/2}\,\sigma_z \otimes \mathbb{I}_m,
\\
 L_3 &= \sqrt{\gamma_m(n_\text{th}+1)}\,\mathbb{I}_q \otimes a,
&
 L_4 &= \sqrt{\gamma_m n_\text{th}}\,\mathbb{I}_q \otimes a^\dagger .
\end{aligned}
\label{eq:lindblad_operators}
\end{equation}
representing qubit relaxation, pure dephasing, and mechanical damping and heating, respectively. In the polaron frame these become
\begin{equation}
\begin{aligned}
\tilde{L}_1 &= \sqrt{\Gamma_1}\,\sigma_- \exp[-2k(a^\dagger - a)],
&
\tilde{L}_2 &= \sqrt{\Gamma_\phi/2}\,\sigma_z,
\\
\tilde{L}_3 &= \sqrt{\gamma_m(n_\text{th}+1)}\,(a - \sigma_z k - \bar{G}),
&
\tilde{L}_4 &= \sqrt{\gamma_m n_\text{th}}\,(a^\dagger - \sigma_z k - \bar{G}) .
\end{aligned}
\label{eq:transformed_lindblad_operators}
\end{equation}
The full master equation in the polaron frame is then
\begin{equation}\label{s40}
\dot{\tilde{\rho}} = -\frac{i}{\hbar}[\mathcal{H}', \tilde{\rho}] + \sum_{j=1}^4 \mathcal{D}[\tilde{L}_j]\tilde{\rho},
\end{equation}
where the Lindblad dissipator is defined as
\begin{equation}\label{s41}
\mathcal{D}[L]\rho = L\rho L^\dagger - \frac{1}{2}\{L^\dagger L, \rho\}.
\end{equation}
To obtain an effective qubit-only description, we assume weak system--bath coupling and invoke the Born approximation,
\begin{equation}
\tilde{\rho} \approx \tilde{\rho}_q \otimes \rho_{m,\text{th}},
\end{equation}
where $\rho_{m,\text{th}}$ is the thermal state of the mechanical oscillator with mean occupation $n_\text{th}$. This factorization is the Born/product-state approximation used to obtain the reduced Markovian rate below. Tracing over the mechanical degrees of freedom leads to the reduced master equation
\begin{equation}
\dot{\tilde{\rho}}_q = -\frac{i}{\hbar}\left[\frac{\hbar}{2}\tilde{\Omega}_q\sigma_z, \tilde{\rho}_q\right] + \sum_j \mathrm{Tr}_m[\mathcal{D}[\tilde{L}_j]\tilde{\rho}].
\end{equation}
Carrying out the traces, the dissipators reduce to
\begin{equation}
\mathcal{D}_{\Gamma_1}[\tilde{\rho}_q] = \Gamma_1\left(\sigma_-\tilde{\rho}_q\sigma_+ - \frac{1}{2}\{\sigma_+\sigma_-, \tilde{\rho}_q\}\right),
\end{equation}
\begin{equation}
\mathcal{D}_{\Gamma_\phi}[\tilde{\rho}_q] = \frac{\Gamma_\phi}{2}(\sigma_z\tilde{\rho}_q\sigma_z - \tilde{\rho}_q),
\end{equation}
and an additional dephasing channel originating from the mechanical bath:
\begin{equation}
\mathcal{D}_\text{mech}[\tilde{\rho}_q] = \gamma_m k^2(2n_\text{th} + 1)(\sigma_z\tilde{\rho}_q\sigma_z - \tilde{\rho}_q).
\end{equation}
Thus, the coefficient multiplying the effective $\sigma_z$ dissipator is
\begin{equation}
\Gamma_\phi' = \frac{\Gamma_\phi}{2} + \gamma_m k^2(2n_\text{th} + 1).
\end{equation}
This effective description shows how mechanical dissipation, when combined with longitudinal coupling, manifests as an additional qubit dephasing channel. Within this factorized reduction, the mechanical contribution to the off-diagonal long-time decay rate is $2\gamma_mk^2(2n_{\rm th}+1)$. In the high-$Q_m$ limit it equals the asymptotic slope $\lim_{t\to\infty}\dot{\chi}_m(t)$ of Eq.~\eqref{eq:mechanical-cumulant}. Accordingly, the finite-time visibility uses $\chi_m(t)$ as its single mechanical-bath factor.
\section{Quantum Fisher Information with Decoherence}
\label{AD}
We now combine the reduced background-qubit dynamics with the stationary finite-time mechanical cumulant to derive the qubit-accessible QFI in the presence of decoherence. In the computational basis $\{|0\rangle, |1\rangle\}$ we write
\begin{equation}
\tilde{\rho}_q = \begin{pmatrix} \rho_{00} & \rho_{01} \\ \rho_{10} & \rho_{11} \end{pmatrix},
\end{equation}
for which the evolution equations implied by the reduced master equation are

\begin{equation}
\begin{aligned}
\dot{\rho}_{00} &= \Gamma_1 \rho_{11}, \\
\dot{\rho}_{11} &= -\Gamma_1 \rho_{11}, \\
\dot{\rho}_{01} &= \left(+i\tilde{\Omega}_q - \frac{\Gamma_1}{2} - 2\Gamma_\phi'\right)\rho_{01}, \\
\dot{\rho}_{10} &= \left(-i\tilde{\Omega}_q - \frac{\Gamma_1}{2} - 2\Gamma_\phi'\right)\rho_{10}.
\end{aligned}
\end{equation}

These equations describe the factorized long-time reduction: relaxation from $|1\rangle$ to $|0\rangle$ occurs at rate $\Gamma_1$, and transverse coherence decays at rate $\Gamma_1/2+2\Gamma_\phi'$. For the finite-time revival calculation, Eq.~\eqref{eq:mechanical-cumulant} replaces the asymptotic mechanical-rate term. To connect back to the laboratory frame, we apply the inverse polaron transformation to the full density matrix,
\begin{equation}
\rho_\text{lab}(t) = \mathcal{T}\tilde{\rho}(t)\mathcal{T}^\dagger,
\end{equation}
and then trace over the mechanical mode:
\begin{equation}
\rho_{q,\text{lab}}(t) = \mathrm{Tr}_m[\mathcal{T}\tilde{\rho}(t)\mathcal{T}^\dagger].
\end{equation}
The diagonal elements are unaffected by this transformation,
\begin{equation}
\rho_{00,\text{lab}}(t) = \tilde{\rho}_{00}(t), \quad \rho_{11,\text{lab}}(t) = \tilde{\rho}_{11}(t).
\end{equation}
For the laboratory-frame product preparation specified in Sec.~\ref{sec:decoherence}, the stationary Gaussian treatment gives the finite-time coherence
\begin{equation}
\rho_{01,\text{lab}}(t)
=\rho_{01,\text{lab}}(0)\,
\exp\!\left[(i\tilde{\Omega}_q-\Gamma_{2,{\rm bg}})t\right]
\mathcal{M}(t),
\end{equation}
with

\begin{equation}
\mathcal{M}(t)
=
\exp[-\chi_m(t)]
\exp[+i\varphi(t)],
\end{equation}

and
\begin{equation}
\varphi(t)
=
4k\bar{G}\sin(\omega_m t).
\end{equation}
This phase retains its high-$Q_m$ form; finite-$Q_m$ phase corrections scale as $\gamma_m t$ and lie below the quoted numerical precision. Thus, the background qubit bath supplies exponential decay, $\chi_m(t)$ supplies the stationary mechanical coherence loss, and $\varphi(t)$ supplies the gravity-dependent phase modulation generated by the longitudinal coupling. For an initial pure qubit state
\begin{equation}
|\psi(0)\rangle
=
\cos(\theta/2)|0\rangle
+
e^{i\phi}\sin(\theta/2)|1\rangle,
\end{equation}
the laboratory-frame density matrix elements can be written explicitly as

\begin{equation}
\begin{aligned}
\rho_{00,\text{lab}}(t)
&=
1-\sin^2\left(\frac{\theta}{2}\right)e^{-\Gamma_1 t},
&
\rho_{11,\text{lab}}(t)
&=
\sin^2\left(\frac{\theta}{2}\right)e^{-\Gamma_1 t},
\\
\rho_{01,\text{lab}}(t)
&=
\frac{1}{2}\sin\theta\,
\exp\left[
+i\big(\Phi(t)-\phi\big)
-\chi_m(t)
-\Gamma_{2,{\rm bg}}t
\right].
\end{aligned}
\label{eq:lab_density_matrix_elements}
\end{equation}

If the initial Ramsey phase is chosen as $\phi=0$, this reduces to the expression used in the main text. The total accumulated phase is
\begin{equation}
\Phi(t)
=
\tilde{\Omega}_q(g)t
+
4k\bar{G}\sin(\omega_m t),
\end{equation}
where
\begin{equation}
\tilde{\Omega}_q(g)
=
\Omega_q-4\omega_m k\bar{G}.
\end{equation}
Equivalently, one may eliminate $\tilde{\Omega}_q(g)$ and write
\begin{equation}
\Phi(t)
=
\Omega_q t
+
4k\bar{G}\big[\sin(\omega_m t)-\omega_m t\big].
\end{equation}
The background transverse-decoherence rate is:
\begin{equation}
\Gamma_{2,{\rm bg}}
=
\frac{\Gamma_1}{2}
+
\Gamma_\phi
=\frac{1}{T_{2R,{\rm bg}}}.
\end{equation}

\revred{We use $\sigma_-=|0\rangle\langle1|$, $\sigma_+=|1\rangle\langle0|$, $\sigma_z=|1\rangle\langle1|-|0\rangle\langle0|$ and $\sigma_y=-i(\sigma_+-\sigma_-)$.} \revred{Thus $r_z=\rho_{11}-\rho_{00}$ and $\rho_{01}=(r_x+i r_y)/2$.} In terms of the Bloch vector $\vec{r}(t)=(r_x(t),r_y(t),r_z(t))$ associated with $\rho_{q,\text{lab}}(t)$, and taking $\phi=0$ for compactness, we have

\begin{align}
\begin{split}
r_x(t)
=&\,
\sin\theta\cos\Phi(t)\,
\exp[-\chi_m(t)-\Gamma_{2,{\rm bg}}t],
\\
r_y(t)
=&\,
\sin\theta\sin\Phi(t)\,
\exp[-\chi_m(t)-\Gamma_{2,{\rm bg}}t],
\\
r_z(t)
=&\,
2\sin^2\left(\frac{\theta}{2}\right)e^{-\Gamma_1 t}-1.
\end{split}
\end{align}

The gravitational field $g$ enters these expressions through
\begin{equation}
\bar{G}
=
\frac{m_{\mathrm{eff}}gz_{\mathrm{zpf}}}{\hbar\omega_m},
\end{equation}
and hence through $\Phi(t)$.

\revred{To compute the QFI for estimating $g$ here, we use the locally fixed mechanical input. In this calculation, $g$ enters the Bloch vector through the phase $\Phi(t)$, while the decoherence factors, populations and decay rates are locally independent of $g$.} \revred{The gravity-dependent loaded preparation is treated separately in Sec.~\ref{sec:loaded-preparation}.} The phase derivative is
\begin{equation}
\frac{\partial\Phi}{\partial g}
=
4k\big[\sin(\omega_m t)-\omega_m t\big]
\frac{m_{\mathrm{eff}}z_{\mathrm{zpf}}}{\hbar\omega_m}.
\end{equation}
For a single qubit with phase-only encoding of $g$, the QFI associated with estimating $g$ from $\rho_{q,\text{lab}}(t)$ is
\begin{align}
\begin{split}
\mathcal{F}_Q(g;t)
=
\sin^2\theta
\exp[&-2\chi_m(t)-2\Gamma_{2,{\rm bg}}t]
\left(\frac{\partial\Phi}{\partial g}\right)^2.
\end{split}
\end{align}
Substituting the explicit derivative gives
\begin{align}\label{s67}
\begin{split}
\mathcal{F}_Q(g;t)
=&\,
\sin^2\theta
\exp[-2\chi_m(t)-2\Gamma_{2,{\rm bg}}t]
\\
&\times
\left[
4k\big(\sin\omega_m t-\omega_m t\big)
\frac{m_{\mathrm{eff}}z_{\mathrm{zpf}}}{\hbar\omega_m}
\right]^2.
\end{split}
\end{align}
Equation~\eqref{s67} displays how decoherence, longitudinal coupling, and interrogation time determine the achievable precision. At short times,
\[
\sin\tau-\tau
=
-\frac{\tau^3}{6}
+O(\tau^5),
\qquad \tau=\omega_m t,
\]
so the leading phase sensitivity scales as $t^3$ and the corresponding qubit-accessible QFI scales as $F_Q^{(q)}\propto t^6$. Finally, we consider a Ramsey-type measurement of $g$ based on projective qubit readout. For a given local-oscillator phase $\phi_\text{LO}$, the associated Classical Fisher Information is
\begin{equation}
\mathcal{F}_C(g;t,\phi_\text{LO})
=
\frac{
r_\perp^2(\partial_g\Phi)^2\sin^2(\Phi+\phi_\text{LO})
}{
1-r_\perp^2\cos^2(\Phi+\phi_\text{LO})
},
\end{equation}
where $ r_\perp=\sqrt{r_x^2+r_y^2}. $

\revred{Setting the measurement phase to $\phi_{\rm LO}=\pi/2-\Phi(g_0;t)$ at a calibrated or preliminary estimate $g_0$, and holding it fixed under $\partial_g$, yields the local maximum at $g=g_0$}
\begin{equation}\label{s69}
\mathcal{F}_C^\text{max}(g;t)
=
r_\perp^2(\partial_g\Phi)^2
=
\mathcal{F}_Q(g;t).
\end{equation}
Thus, an ideal Ramsey quadrature measurement saturates the qubit QFI for this phase-encoding model. Finite readout fidelity may be included separately through an effective reduction factor such as
\begin{equation}
\mathcal{F}_{\rm eff}(g;t)
\simeq
(2F_r-1)^2\mathcal{F}_Q(g;t).
\end{equation}
\bigskip
\section{Effects of longitudinal coupling \texorpdfstring{$k$}{k}}
\label{sec:k-effects}
We evaluate the illustrative values $k=0.1$ and $0.3$. The coupling is constrained by Eq.~\eqref{eq:coupling-map}; each sweep point therefore uses
\begin{equation}
D_m(k)=\frac{2f_mk}{\delta\varphi_{\rm zpf}}.
\end{equation}
The blue curves retain the ideal $k^2$ QFI scaling. The red curves simultaneously include the corresponding increase of mechanical-loop flux-noise dephasing through $D_m(k)$, together with the other visibility factors in Eq.~\eqref{eq:visibility-model}. These panels are diagnostic sweeps; the selected operating values remain those in Table~\ref{tab:scenarios_combined}.
\begin{figure*}[!t]
\centering
\vspace{6pt}
\begin{minipage}[t]{0.48\textwidth}
  \centering
  \includegraphics[width=0.5\linewidth,height=1.15in]{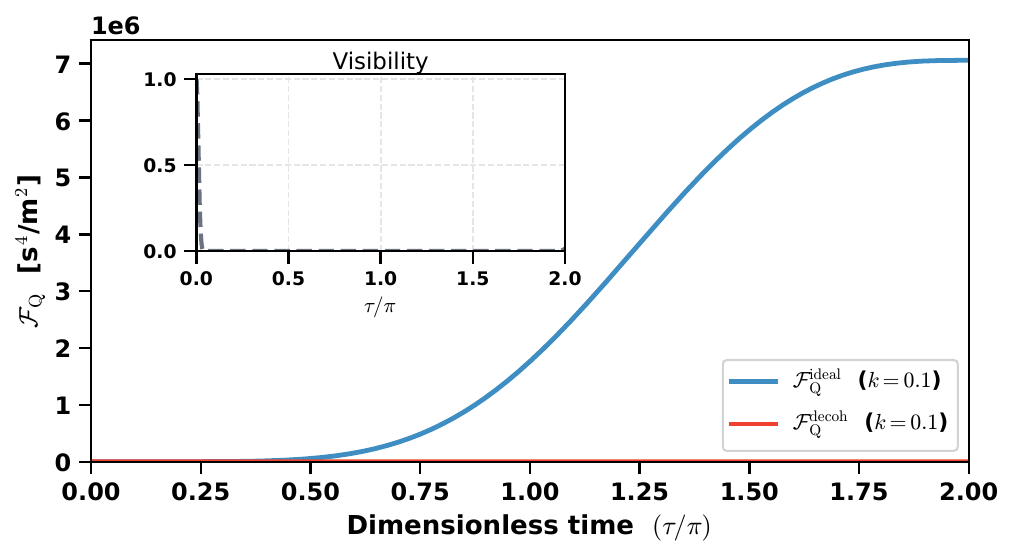}\hfill
  \includegraphics[width=0.5\linewidth,height=1.15in]{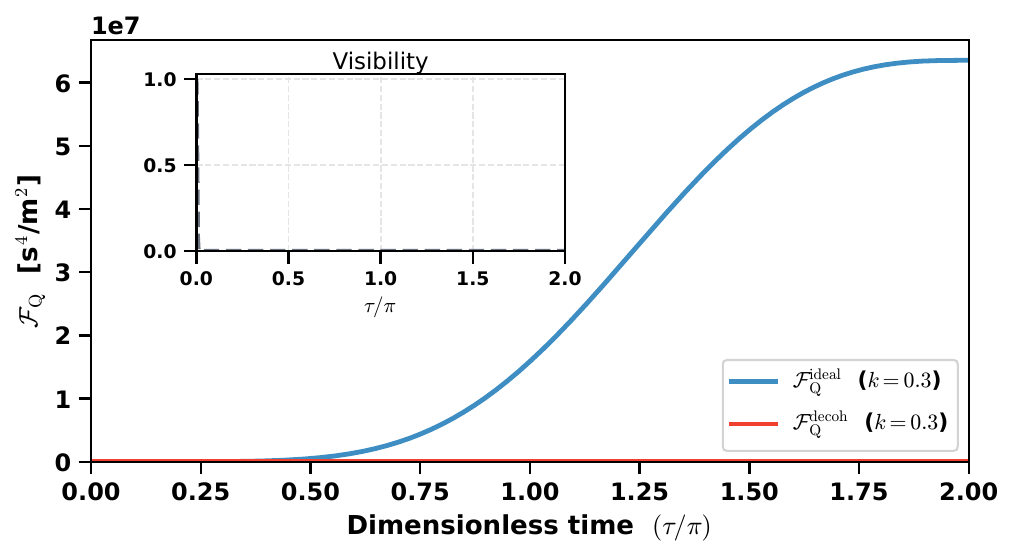}
  \textbf{Scenario I (component-anchored benchmark)}\\[4pt]
\end{minipage}\hfill
\begin{minipage}[t]{0.48\textwidth}
  \centering
  \includegraphics[width=0.5\linewidth,height=1.15in]{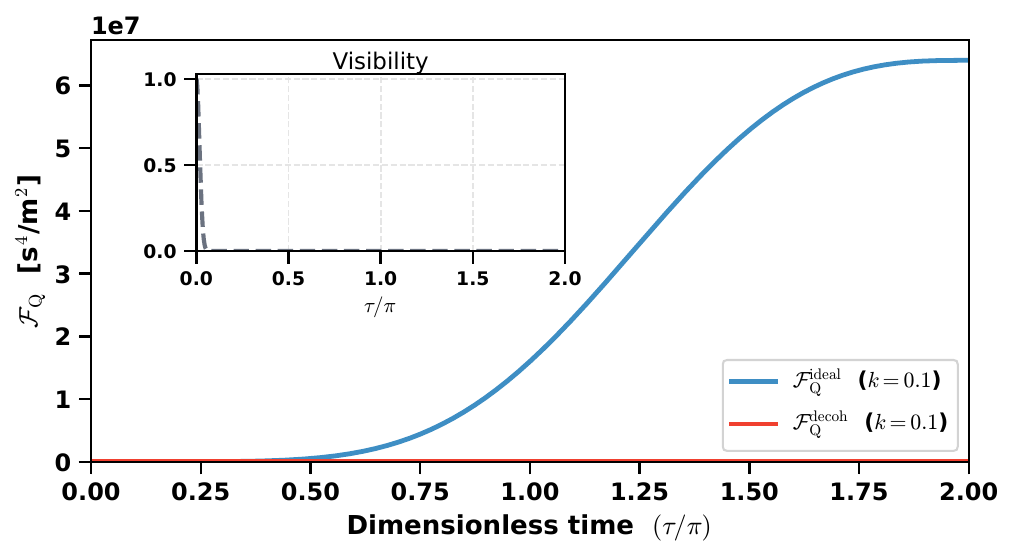}\hfill
  \includegraphics[width=0.5\linewidth,height=1.15in]{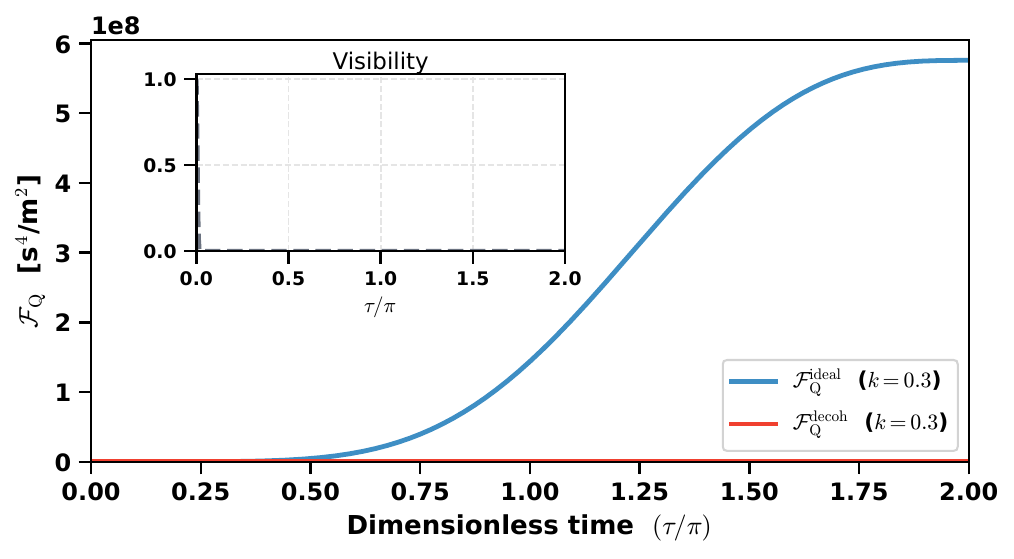}
  \textbf{Scenario II (long-term integration target)}\\[4pt]
\end{minipage}
\vspace{6pt}
\caption{\justifying \revred{Circuit-constrained comparison of longitudinal Pauli couplings $k=0.1$ and $k=0.3$ for both configurations, using the locally fixed mechanical preparation and phase slope of Eq.~\eqref{eq:FQ_full}.} The panels show QFI versus $\tau/\pi$. Blue curves are ideal; red curves use $D_m(k)=2f_mk/\delta\varphi_{\rm zpf}$ and the visibility in Eq.~\eqref{eq:visibility-model}. Insets show the corresponding visibility. {On this linear scale the red curves are indistinguishable from zero: with $D_m(k)$, the first-revival coherence amplitude is $W(t_1)\approx1.2\times10^{-2}$ for Scenario~I at $k=0.1$ and smaller in the remaining panels.}}
\label{fig:comparisonk}
\end{figure*}
Figure~\ref{fig:comparisonk} shows why $k$ cannot be varied independently of the circuit response: increasing $k$ raises the ideal signal and the same-loop flux-noise susceptibility together.
\end{widetext}
\end{document}